\documentclass[
a4paper,
prb,
bibnotes,
showpacs,
twocolumn,
superscriptaddress,
nofootinbib,
]{revtex4-1}
\usepackage{graphicx}
\usepackage{amssymb,amsmath,color}
\usepackage{amsmath}
\usepackage[colorlinks=true,citecolor=blue]{hyperref}
\newcommand{\Ki}{P^i} 
\newcommand{\CT}{{T}_C} 
\newcommand{\rrmp}{m_{\rm p}} 
\newcommand{\rrme}{m_{\rm e}} 

\newcommand{\kappaa}{G_h}
\newcommand{\ZT}{Z_eT}
\newcommand{\tauu}{t'}
\newcommand{\schrodinger}{Schr\"{o}dinger\,\,}
\newcommand{\hc}{ {\rm H.c.}} 
\newcommand{\tr}{\mbox{Tr}}

\newcommand{\rrho}{\tilde{\rho}} 
\newcommand{\rme}{\rm e} %
\newcommand{\rmpp}{\rm p} %
\newcommand{\rmep}{\rm ep} %
\newcommand{\Hs}{H_S} 
\newcommand{\Ho}{H_0} 
\newcommand{\Hse}{V_{\rm e}} 
\newcommand{\iV}{V_I} 
\newcommand{\frho}{\rho} 
\newcommand{\irho}{\rho_I} 
\newcommand{\ir}{\tilde{\rho}_I} 
\newcommand{\U}[2]{e^{-i#1 #2/\hbar}} 
\newcommand{\Ud}[2]{e^{i#1 #2/\hbar}} 
\newcommand{\E}{B}  
\newcommand{\Ej}{\mathcal{B}} 
\newcommand{\iS}{S_I} 
\newcommand{\iE}{\E_I} 

\newcommand{\ob}{\mathcal{J}}  
\newcommand{\io}{\ob_I} 

\begin{document}
\title{Effects of electron-phonon interaction on thermal and electrical transport through molecular nano-conductors}
\author{Jing-Tao L\"u}
\email{jtlu@hust.edu.cn}
\affiliation{School of Physics, Huazhong University of Science and Technology, 430074 Wuhan, People's Republic of China}
\author{Hangbo Zhou}
\affiliation{Department of Physics and Center for Computational Science and Engineering, National University of Singapore, 117551 Singapore, Republic of Singapore}
\affiliation{NUS Graduate School for Integrative Sciences and Engineering, National University of Singapore,  117456 Singapore, Republic of Singapore}
\author{Jin-Wu Jiang}
\affiliation{Shanghai Institute of Applied Mathematics and Mechanics, Shanghai Key Laboratory of Mechanics in Energy Engineering, Shanghai University,  200072 Shanghai, People's Republic of China}
\author{Jian-Sheng Wang}
\affiliation{Department of Physics and Center for Computational Science and Engineering, National University of Singapore, 117551 Singapore, Republic of Singapore}
\date{January 26, 2015}
\begin{abstract}
The topic of this review is the effects of electron-phonon interaction (EPI) on
the transport properties of molecular nano-conductors. A nano-conductor
connects to two electron leads and two phonon leads, possibly at different
temperatures or chemical potentials. The EPI appears only in the nano-conductor.
We focus on its effects on charge and energy transport.  We introduce three
approaches.  For weak EPI, we use
the nonequilibrium Green's function method to treat it perturbatively. We
derive the expressions for the charge and heat currents. For weak system-lead
couplings, we use the quantum master equation approach. In both cases, we use a
simple single level model to study the effects of EPI on the system's
thermoelectric transport properties.  It is also interesting to
look at the effect of currents on the dynamics of the phonon system. For this,
we derive a semi-classical generalized Langevin equation to describe the
nano-conductor's atomic dynamics, taking the nonequilibrium electron system,
as well as the rest of the atomic degrees of freedom as effective baths. We show
simple applications of this approach to the problem of energy transfer between
electrons and phonons.
\end{abstract}
\maketitle

\section{Introduction}
Electron-phonon interaction (EPI) is one of the most important many-body
interactions in condensed-matter and molecular systems\footnote{We do not distinguish
vibrations in isolated molecules and phonons in periodic lattices.}, responsible for a
variety of phenomena, from electrical, thermal conduction, superconductivity to
Raman scattering, polaron formation, just to list a few\cite{born-huang,born-oppenheimer,vanLeeuwen04,Ziman-ep,grimvall,shamziman63,hubac08}. 
Its effects on the electrical, thermal, and optical properties of 
bulk semiconductors and metals have been intensively studied along with the development
of many-body theories and experimental techniques. Recent
advances in experimental fabrication of meso- and nano-scopic structures have
generated tremendous efforts in understanding the effects of EPI on transport
properties of reduced-dimensional systems\cite{GaRaNi.2007,HoBoNeSaToFi06,Ta.2006}.  

Of special interest are current-induced forces and Joule heating in
low-dimensional systems, especially in molecular
nano-conductors\cite{To.1998,ToHoSu.2000,DiChTo.04,DuMcTo.2009,ToDuMc.2010,JMP.2010,luprb12,BoKuEgVo.2011,Abedi10,Abedi13,StReHo98,ness_quantum_1999,Agrait2002,SmUnRu.2004,SeRoGu.2005,ViCuPaHa05,DeLaVega2005,Paulsson2005,HuangZ2006nl,TsutsuiM2007apl,Lu07,Asai08,Lu08JPCM,IoffeZ2008nn,WuBH2009jpcm,YuYJ2011apl,WardDR2011nn,ShekharS2011apl,Hartle11,ShekharS2012carbon,Hutzen12,Simine13PCCP,Simine13PI,Wilner13,bergfield_forty_2013,Kaasbjerg13,WangQiang13,Simine14}.
On the one hand, the electrical transport signature of EPI is an invaluable
spectroscopic tool to study the structural information of molecular
nano-conductors\cite{StReHo98,Agrait2002}.  On the other hand, these
processes are crucial in maintaining the stability of these conductors\cite{SmUnRu.2004}, relevant to the continuous scaling down of modern
electronic devices. Different theoretical approaches have been developed to
study these problems, in many cases separately. Recently, it was realized that
non-conservative nature of current-induced forces provides an alternative,
deterministic way of energy transfer between electrons and phonons, or more
generally atomic motions\cite{DuMcTo.2009}.  It is fundamentally different from
the stochastic Joule heating. These advances have motivated the development of
methods treating current-induced forces and Joule heating on the same footing\cite{JMP.2010,BoKuEgVo.2011,luprb12,Lu08JPCM}. 

Equally significantly, there has been an increasing interest in the thermoelectric
properties of low dimensional systems\cite{dresselhaus_new_2007,DuDi11,li_colloquium:_2012,Yang12Rev,Yang12}. A
starting point of the theoretical treatment is to ignore the effect of EPI, and
study the transport of electrons and phonons separately. But how important 
the effect of EPI is is a pertinent question, on which much of recent work is
devoted
to\cite{Leijnse2010,Hsu2010,sergueev_efficiency_2011,entin-wohlman_three-terminal_2010,Zhou2015}.
Here, we will look at this problem using the various approaches we have developed.

EPI is a genuine many-body interaction, the exact treatment of which is
challenging, if possible at all. One natural approach is to perform perturbation
calculation over a certain small parameter. In the most common multi-probe
transport setup (see Fig.~\ref{fig:sys} and Sec.~\ref{sec:epi}), this small
parameter can be chosen according to the strength of EPI. This strength can be
roughly characterized by the ratio between two time scales: the first one corresponds
to the phonon period, and the second one corresponds to the electron
dwell time\cite{buttiker83} in the nano-conductor. If the time electrons spend in the
nano-conductor is much shorter than the phonon period, the system is in the weak EPI
regime. The small parameter is the EPI matrix. In the other limit, the coupling
of the nano-conductor to electrodes is the small parameter, over which one can
perform the perturbation expansion.

In this review, we summarize our own effort in  developing and/or utilizing
different theoretical approaches to study the aforementioned problems in
different parameter regimes. We discuss some relevant results when possible, 
but we make no effort on reviewing all of them considering the huge amount of literature.
The paper is organized as follows: In
Sec.~\ref{sec:epi}, we give a brief introduction of the EPI problem starting
from the Born-Oppenheimer approximation. We then introduce the system setup
and Hamiltonian we use in this paper. In Sec.~\ref{sec:negf}, we briefly
summarize our use of the nonequilibrium Green's function (NEGF) method to 
study electron, phonon transport and their interaction perturbatively. 
We consider several applications of the method. The first
one is the effects of EPI on the thermoelectric transport coefficients in a
single level model. The second one is the heat transport between electrons and
phonons due to EPI.  
The use of simple models enables us to approach the problems
semi-analytically. The last example is a numerical study of the Joule heating 
and phonon-drag effect in carbon nanotubes. In Sec.~\ref{sec:qme}, we consider the case of strong EPI
using the quantum master equation (QME) approach. After reviewing the earlier work, the same thermoelectric
transport model is re-visited focusing on how the strength of EPI affects the
results. In Sec.~\ref{sec:gle}, we focus on the current-induced dynamics. Based
on the Feynman-Vernon influence functional approach, we derive the
semi-classical Langevin equation, taking into account the equilibrium phonon
and nonequilibrium electron baths. The final section is our conclusion and
remarks.

\section{Born-Oppenheimer approximation and electron-phonon interactions}
\label{sec:epi}
To discuss the meaning and formulation of the electron and phonon systems and their
mutual interaction, we need to start from the Born-Oppenheimer 
approximation\cite{born-oppenheimer,born-huang}.  Consider an electron-ion system
with a total Hamiltonian $H = \Ki + H^e$, where
$\Ki$ is the kinetic energy operator for the ions, and $H^e = P^e + U$ is 
electron Hamiltonian with kinetic energy of the electrons, $P^e$, and
potential energy $U = U^{ee} + U^{ei} + U^{ii}$, which includes
the Coulomb interactions among the electrons and ions.  
Since the ions are much
heavier than the electrons, one can treat the ion kinetic energy term as a small perturbation
with the expansion parameter\cite{born-huang}
\begin{equation}
	\left( \frac{m_{\rm e}}{m_{\rm p}} \right)^{1/4},
\end{equation}
where $\rrme$ is the mass of an electron and $\rrmp$ mass of an ion (assuming all have
the same mass).  If the ions are considered infinitely heavy, the ions will not move 
and the electron wavefunctions satisfy
\begin{equation}
\label{Hpsi-adiabatic}
H^e \phi_\alpha(x;R) = E^e_\alpha(R) \phi_\alpha(x;R),
\end{equation}
where $x$ represents the set of all coordinates of the electrons, $R$ the positions
of all the ions, and $\alpha$ is the electronic state quantum number.  The eigen-functions
and the eigenvalues depend on $R$ parametrically.  

We assume an orthonormal set $\{\phi_\alpha\}$ that satisfies Eq.~(\ref{Hpsi-adiabatic}) has been obtained.
To take into account the effects of the ions, we consider a trial full wavefunction in a 
factored form
\begin{equation}
\label{BO-basis}
\Psi(x,R) = \phi_\alpha(x;R) \chi_{\beta;\alpha}(R) = |\alpha\beta\rangle,
\end{equation}
and consider the variational solution\cite{vanLeeuwen04} of the full Hamiltonian, 
${\rm min}_{\chi} 
\langle \Psi | H | \Psi \rangle $, subject to the normalization $\langle
\chi | \chi \rangle = 1$.   This variational approach is equivalent to omitting
the off-diagonal elements (which is the Born-Oppenheimer approximation,
see Ref.~\onlinecite{born-huang}, App.~VIII), giving an equation for the ions
\begin{eqnarray}
\Bigl(\Ki + E^e_\alpha(R) + 
 \langle \phi_\alpha | \Ki | \phi_\alpha \rangle \qquad\qquad\qquad  \nonumber \\
\qquad\qquad -
\frac{\hbar^2}{m_{\rm p}} \langle \phi_\alpha | \nabla_R | \phi_\alpha \rangle \cdot \nabla_R 
\Bigr) \chi  = E \chi,
\end{eqnarray}
where $\langle \cdots \rangle$ means the $x$-dependence is integrated
out but still $R$-dependent;  $\nabla_R$ is a multi-dimensional gradient operator
with respect to $R$.  Since the left-hand side depends on the electronic quantum
number $\alpha$, the full eigen-energy $E$ and functions also depend on $\alpha$
parametrically, e.g., we may write $E_{\beta;\alpha}$. 

If we assume that the electrons are in its instantaneous ground state, the ions
move in a potential surface generated by the electrons.  There are no explicit
electron-phonon interaction (EPI) terms.  To account for the EPI, we
need to go back to the basis, Eq.~(\ref{BO-basis}), and consider the matrix
elements
\begin{equation}
\langle \alpha \beta | H |\alpha' \beta' \rangle.
\end{equation}
The off-diagonal terms are interpreted as the EPI\cite{Ziman-ep,grimvall}, which are small.  If the off-diagonals are
omitted, the electrons stay in a given quantum state $\alpha$.  The off-diagonal terms
describe the scattering of the electrons to different state $\alpha'$.  If ion displacements are small, the 
most important contribution is from the linear term in the displacement
\begin{equation}
\label{BO-off-diagonals}
- \frac{\hbar^2}{\rrmp} \langle \phi_\alpha \chi_{\beta;\alpha} | \nabla_R | \phi_{\alpha'} \rangle \cdot \nabla_R | \chi_{\beta';\alpha'} \rangle,\quad
 (\alpha,\beta) \neq ( \alpha',\beta').
\end{equation}
These off-diagonal matrix elements can be used, e.g., in a Fermi-Golden rule
calculation of scattering processes.  However, the identity (in the sense of 
effective Hamiltonians) of the electrons and phonons and their mutual interaction
are not at all clear.   Although EPI
plays major role in many physical processes\cite{shamziman63}, such as electronic transport and superconductivity, its conceptual foundation is still not very solid. 
Within the Born-Oppenheimer scheme, it is not clear at all
how to transform the original Hamiltonian $H$ into a form of an electron system
and independent phonon system and their interaction unambiguously.  The 
problem is related to the fact that in deriving the phonon Hamiltonian (the potential
surfaces), the effect of electrons is already used.  Thus, putting the electrons
back amounts to double counting, see Refs.~\onlinecite{vanLeeuwen04,hubac08} for some of the modern treatments.   

Instead of pursuing a self-consistent theory of EPI from the Born-Oppenheimer
approximation, here in this review, and also in many of the practical
applications\cite{deegan72,ashkenazi81,cappelluti06,FrPaBr.2007,giustino07}, we
adopt a phenomenological point of view, and use the model Hamiltonians as given
below in Eqs.~(\ref{pheno-Htot}) and (\ref{pheno-Hep}).   Focusing only the
term linear in the displacements away from the equilibrium positions of the
ions, we can think of the single electron Hamiltonian $H_{\rm e}$ below having
a $R$-dependence. Taylor expanding it, $R = R^0 + u/\sqrt{\rrmp}$, we obtain
\begin{equation}
\label{Mijk}
M_{ij}^k = \frac{1}{\sqrt{\rrmp}} \, \langle i | \frac{\partial H_{\rm e}(R)}{\partial R_k} | j\rangle,
\end{equation}
where $ | j\rangle$ is the single particle state when ionic system is in equilibrium
position $R^0$. The extra factor of square root of ion mass $\rrmp$ is because of our convention of displacement variable $u$. 
This form of interaction is intuitively understandable and originally proposed by
Bloch\cite{bloch28}.   In Chap.~4 of Ref.~\onlinecite{grimvall}, a derivative
from Eq.~(\ref{BO-off-diagonals}) to (\ref{Mijk}) is given, but the reasoning
does not seem to be rigorous. 

\begin{figure}[!htbp]
	\includegraphics[scale=0.4]{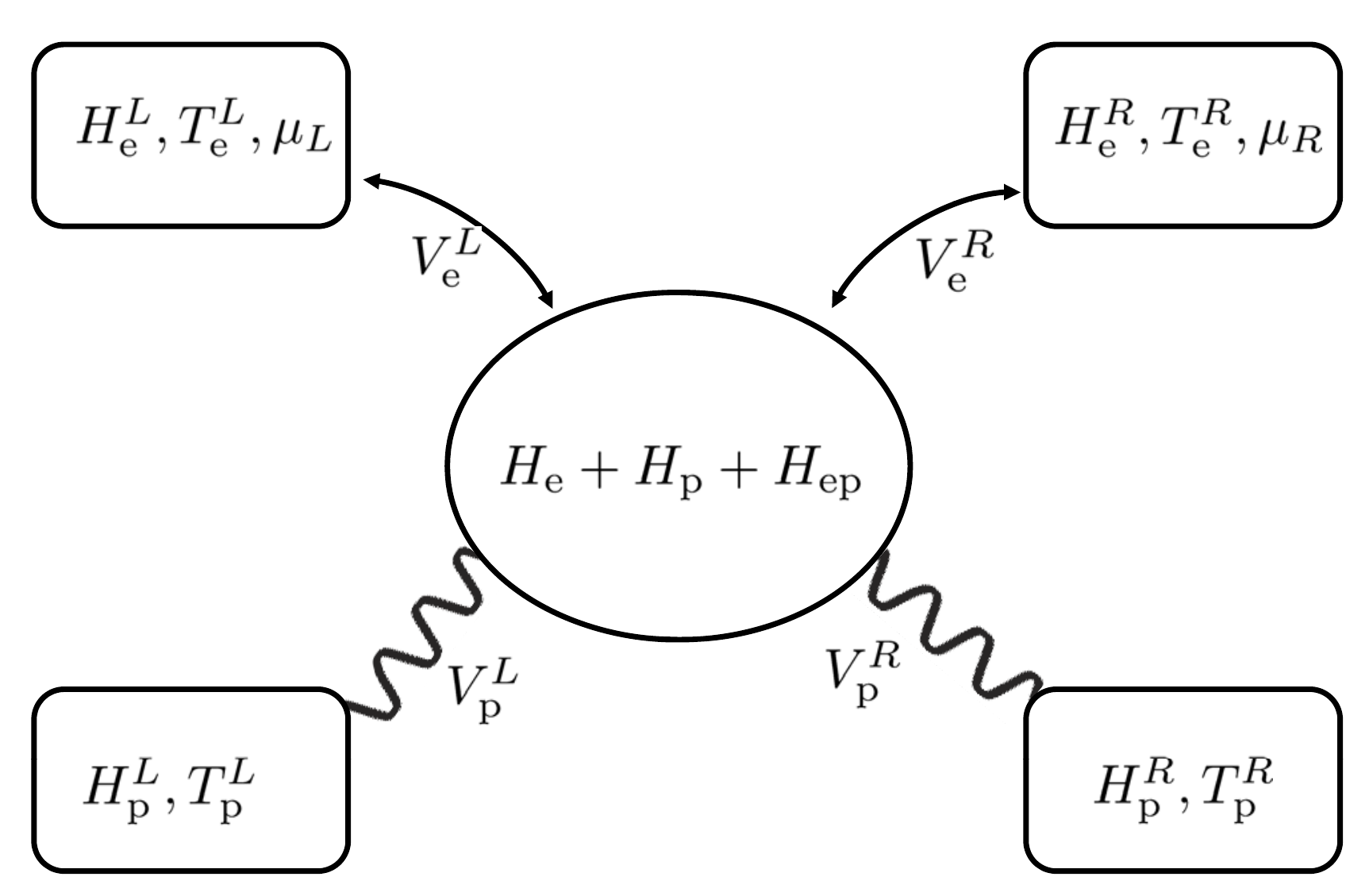}
	\caption{Model system considered in this review. The center device, including both electrons, phonons, and their interactions, is coupled with two electron and two phonon leads. Each electron lead is characterized by its chemical potential $\mu_\alpha$ and temperature $T^\alpha_{\rm e}$, and each phonon lead by temperature $T^\alpha_{\rm p}$.} 
	\label{fig:sys}
\end{figure}

Thus, our starting point of a derivation is a tight-binding Hamiltonian for
the electrons, harmonic couplings for the phonons and a standard
EPI term.  They are taken as given and exact.
The charge redistributions and self-consistency for the
electrons are not part of the discussion.  Symbolically, the total
many-body Hamiltonian is given as
\begin{equation}
\label{pheno-Htot}
H_{{\rm tot}} = H^0_{{\rm e}} + H_{{\rm p}} + H_{{\rm ep}},
\end{equation}
where the electron part is $H^0_{{\rm e}} = c^\dagger H c$, the phonon
part $H_{{\rm p}} = \frac{1}{2}( p^T p + u^T K u) +V_n(u^C)$.  
The variable $u$ is mass normalized, $u_j = \sqrt{m_{j}} (R_j - R_j^0)$.
Because of this, the conjugate momentum is $p = \dot{u}$. $V_n$ is the 
nonlinear force contribution. $c$ is
a column vector of the electron annihilation operators, which we can separate
into three regions, the left, center, and right, $c=(c^L, d,
c^R)^T$, $T$ stands for matrix transpose.  Similarly $u=(u^L, u^C,
u^R)^T$.  Accordingly, the matrices $H$ and $K$ are partitioned into
nine regions (submatrices), e.g.,
\begin{equation}
H = \left( \begin{array}{ccc}
    H^L & H^{LC} & 0 \\
    H^{CL} & H^C & H^{CR} \\
    0 & H^{RC} & H^R 
    \end{array} \right),
\end{equation}
such that $H^0_{\rm e} = H_{\rm e}^L +H_{\rm e}^R+H_{\rm e}^C+V_{\rm e}$,
with $V_{\rm e} = V_{\rm e}^L+V_{\rm e}^R$, $V_{\rm e}^L = {c^L}^\dagger H^{LC} d +{\rm H.c.}$.
Note that we assume no interaction between the left and right leads (See
Ref.~\onlinecite{LiAgWa12} for transport when there is a lead-lead coupling).
We do a similar partitioning for $K$ using the notation of
Ref.~\onlinecite{WaWaLu07}. The EPI takes the form
\begin{equation}
\label{pheno-Hep}
H_{{\rm ep}}(d,u^C) = \sum_{ijk} M_{ij}^k u^C_k d_i^\dagger d_j = 
\sum_{k} u^C_k d^\dagger M^k d.
\end{equation}
We assume that the EPI appears only in the
central region. A schematic representation of the system setup is shown
in Fig.~\ref{fig:sys}.

The separation of the electron and phonon leads makes the theoretical
development easier. In reality, they could either be physically separated, or
built into one.  For example, one electrode could serve both as an
electron and a phonon lead,  but we assume that we have independent control
over their temperatures $T^\alpha_{\rm e}$ and $T^\alpha_{\rm p}$, $\alpha=L,R$.

\section{Weak EPI regime: Nonequilibrium Green's function method}
\label{sec:negf}
\subsection{Theory}
We first consider the case where EPI is weak, so that we can perform a
perturbation expansion over the interaction matrix $M$. In order to do so, we
use the NEGF method. Detailed introduction is given in our previous
work\cite{WaWaLu07,WaAgLiTh14,Lu07,Lu08}.  This section can be considered as an
application of the general approach developed in
Refs.~\onlinecite{WaWaLu07}-\onlinecite{WaAgLiTh14} to the EPI problem.  We use
similar notations therein, and only give a brief outline of the approach here.

We denote the electron device Green's function without and with EPI by $G_0$
and $G$, the corresponding phonon Green's
functions by $D_0$ and $D$, and the lead Green's functions without coupling to the center as $g_\alpha$ and $d_\alpha$, respectively. The couplings of the device with the
leads and that between the electrons and phonons are described by self-energies,
with $\Sigma$ and $\Pi$ representing that of electron and phonon, respectively.
For example, we define the time-ordered electron Green's function including EPI 
on the Keldysh contour [Fig.~\ref{fig-Cpath} (b)]
\begin{equation}
	G_{ij}(\tau,\tau') = -\frac{i}{\hbar}\langle \CT c_i(\tau) c^\dagger_j (\tau')\rangle.
	\label{eq:cong}
\end{equation}
Here, $\tau/\tau'$ is time on the contour, and $i/j$ is index of the electronic states.
The contour time order operator $T_C$ puts the operators later in the contour to the left.
The average $\langle \cdot \rangle$ is with respect to the density matrix of the full
Hamiltonian.

The contour ordered Green's function can be divided into different groups according to the spatial
position of $i/j$, similar to the Hamiltonian. The most interesting one is 
$G^C$, where $i$ and $j$ are both at the center device region. 
At the same time, it can be written as a $2\times 2$ matrix in time space
\begin{equation}
	G(\tau_i,\tau_j) = \left( \begin{array}{cc} G^t(t_i,t_j) & G^< (t_i,t_j) \\ G^> (t_i,t_j) & G^{\bar{t}}(t_i,t_j)\end{array} \right),
	\label{eq:g22}
\end{equation}
with $G^t$, $G^{\bar{t}}$, $G^>$, $G^<$ the time-ordered, anti-time-ordered,
greater and lesser Green's functions. The retarded and advanced Green's functions
are obtained from them, i.e., $G^r = G^t - G^<$, and $G^a = G^< - G^{\bar{t}}$.
For the definition and relations among these Green's functions, we refer to
the book by Haug and Jauho\cite{HaugJauho}, and our previous
publications\cite{Lu07,WaWaLu07,WaAgLiTh14}.

To calculate
the Green's functions, we use a process of two-step adiabatic switch on.
We start from the decoupled system and leads. Each of the electron and phonon leads
is at its own equilibrium state, characterized by the temperature $T^\alpha$ and/or
chemical potential $\mu_\alpha$. The corresponding equilibrium Green's functions can
thus be defined according to the equilibrium canonical distribution.
The initial state of the system is arbitrary and not important in most cases (e.g., for steady state).

At the first step, we switch on the interaction of the center Hamiltonian 
with the electron and phonon leads. We wait until the electron and phonon subsystem reaches
their own nonequilibrium steady state, since the temperature and/or chemical potential of each lead can be different. 
The two subsystems are quadratic and exactly solvable, and we get the non-interacting center Green's
functions $G_0$ and $D_0$ from the Dyson equation (we omit the superscript $C$)
\begin{eqnarray}
	\label{eq:dyson00}
	G_0(1,2) &=& g_C(1,2) + g_C(1,3)\Sigma_{\rm b}(3,4) G_0(4,2),\\
	\label{eq:dyson01}
	D_0(1,2) &=& d_C(1,2) + d_C(1,3)\Pi_{\rm b}(3,4) D_0(4,2).
\end{eqnarray}
Here, we have used a single number to represent the matrix indices and contour
time arguments, i.e., $G_0(1,2) = {G_0}_{j_1j_2}(\tau_1,\tau_2)$. Summation or integration
over repeated indices is assumed. $g_C$ ($d_C$) is
the center electron (phonon) Green's function without coupling to the $L$ and
$R$ leads. The self-energy $\Sigma_{\rm b}=\Sigma_L+\Sigma_R$ includes
contributions from $L$ and $R$, with $\Sigma_\alpha (1,2) = H^{C\alpha} g_\alpha(1,2)
H^{\alpha C}$; similarly for $\Pi_{\rm b}$.

At the second step, we adiabatically switch on the EPI in the center.
We perform a perturbation expansion over the interaction Hamiltonian $H_{\rmep}$, using
Feynman diagramatics. The interacting Green's functions are expressed using similar Dyson
equations as Eqs.~(\ref{eq:dyson00}-\ref{eq:dyson01}),
\begin{eqnarray}
	G(1,2) &=& G_0(1,2) + G_0(1,3)\Sigma_{\rm ep}(3,4) G(4,2),\\
	D(1,2) &=& D_0(1,2) + D_0(1,3)\Pi_{\rm ep}(3,4) D(4,2).
	\label{eq:dyson}
\end{eqnarray}
Here, $\Sigma_{\rm ep}$ and $\Pi_{\rm ep}$ are electron and phonon self-energies due to EPI.
Using Eq.~(\ref{eq:g22}), at steady state, we can get the following
useful relations in energy/frequency domain
\begin{eqnarray}
	G^r(\varepsilon) &=& \left[(\varepsilon+i 0^+) I - H^C - \Sigma^r_{\rm tot}(\varepsilon)\right]^{-1},\label{eq:grr}\\
	D^r(\omega) &=& \left[(\omega+i 0^+)^2 I - K^C - \Pi^r_{\rm tot}(\omega)\right]^{-1},\\
	\Sigma^r_{\rm tot}(\varepsilon) &=& \Sigma^r_{\rm b}(\varepsilon)+\Sigma^r_{\rm ep}(\varepsilon),\\
	\label{eq:pitot}\Pi^r_{\rm tot}(\omega) &=& \Pi^r_{\rm b}(\omega)+\Pi^r_{\rm ep}(\omega).
	\label{}
\end{eqnarray}
We use $\varepsilon$ for the energy of electron and $\omega$ for the angular frequency
of phonon, respectively, and $I$ is the identity matrix.  
To get an expression for the current, we also need the greater and lesser version
of the Green's functions\cite{HaugJauho}
\begin{eqnarray}
	G^{>,<}(\varepsilon) &=& G^r(\varepsilon) \Sigma_{\rm tot}^{>,<}(\varepsilon)G^a(\varepsilon),\\
	D^{>,<}(\omega) &=& D^r(\omega) \Pi_{\rm tot}^{>,<}(\omega)D^a(\omega).
	\label{eq:dgl}
\end{eqnarray}
The electrical current ($I_{\rm e}$) is expressed as the change rate of the electron
number in one of the leads ($N^\alpha$) times the charge of electron ($-e$). For example, 
\begin{eqnarray}
	I_{\rm e} &=& -e\langle \frac{d N^{L}(t)}{dt}\rangle  \nonumber\\
	&=& -\frac{2e}{\hbar}{\rm Im} {\rm Tr}\left[V^{LC}_{\rme} \langle {c^L}^\dagger(t) d(t) \rangle  \right]\nonumber\\
	&=& 2e {\rm Re} {\rm Tr}\left[ V^{LC}_{\rme} G^{<,CL}(t=0) \right].
	\label{eq:Ia}
\end{eqnarray}
It can be expressed by the Green's function of the center region and the lead self-energies\cite{MeirWingreen,Jauho94,HaugJauho},
\begin{eqnarray}
	I_{\rm e} = \frac{e}{\hbar}\int_{-\infty}^{+\infty}\frac{d\varepsilon}{2\pi} {\rm Tr}\left[ G^>\Sigma_L^<-G^<\Sigma_L^>\right].
	\label{}
\end{eqnarray}
Similarly for the heat current carried by electrons ($I_{\rm h}$) and phonons ($I_{\rm p}$)
\begin{eqnarray}
	I_{\rm h} &=&\frac{1}{\hbar}\int_{-\infty}^{+\infty}\frac{d\varepsilon}{2\pi}(\varepsilon-\mu_L) {\rm Tr}\left[ G^>\Sigma_L^<-G^<\Sigma_L^>\right],\\
	\label{eq:iph}
	I_{\rm p}&=&-\int_{-\infty}^{+\infty}\frac{d\omega}{4\pi}\hbar\omega {\rm Tr}\left[ D^>\Pi^<_L-D^<\Pi^>_L\right].  \label{}
\end{eqnarray}
We have defined the positive current direction as electrons going from the lead to
the center. We dropped the argument of the Green's functions for simplicity.
We ignore the spin degrees of freedom, since it is not relevant here. Currents
out of the right lead are obtained by replacing index $L$ by $R$. 
One can symmetrize the expressions based on energy and charge
conservation.

\begin{figure}[!htbp]
	\includegraphics[scale=0.4]{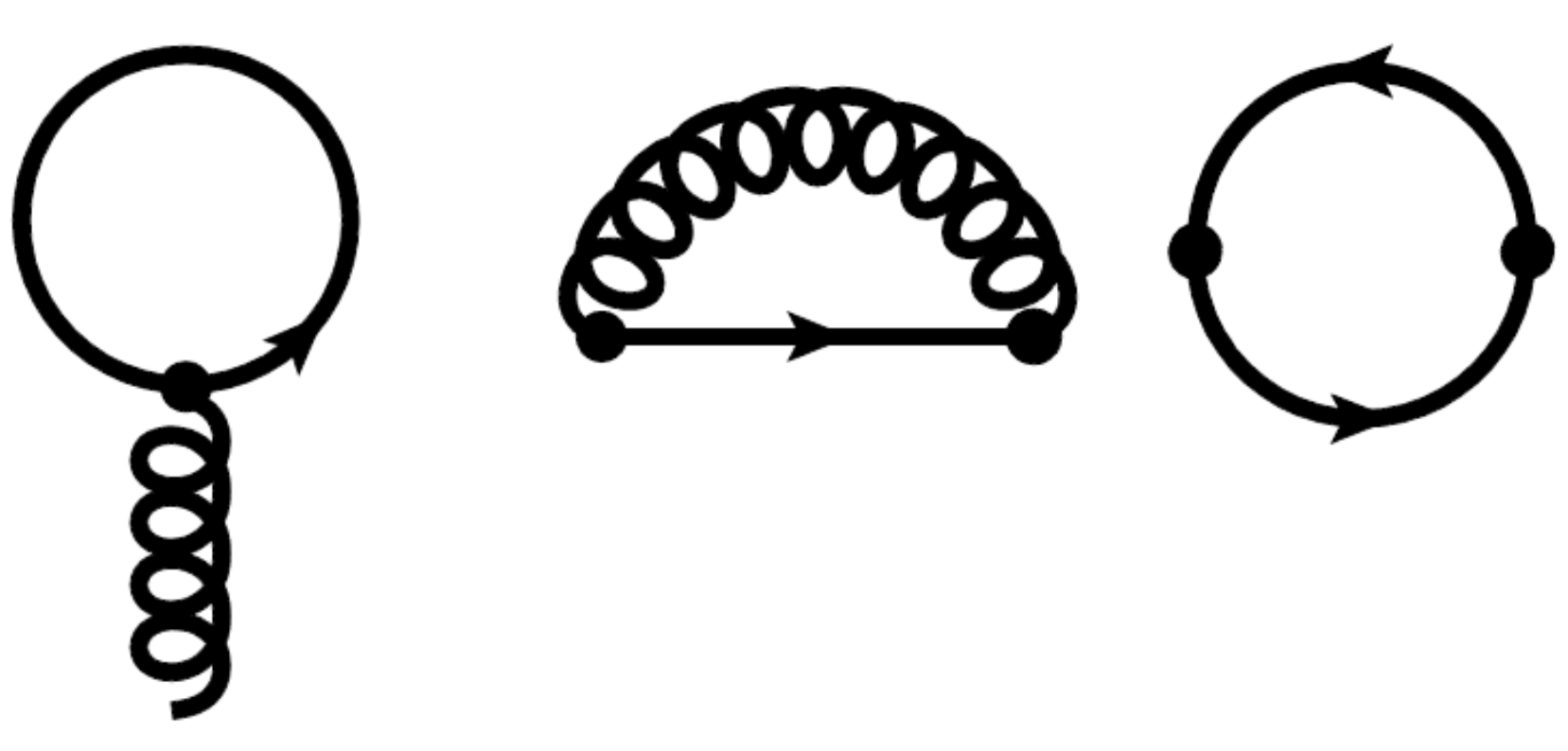}
	\caption{Feynman diagrams due to electron-phonon interaction. The first two are Hartree
	and Fock diagram for electrons, and the last one is the polarization bubble for
	phonons. The expressions of these diagrams can be found in Eqs.~(\ref{eq:eselg}-\ref{eq:ser}).} 
	\label{fig:fey}
\end{figure}

The set of coupled equations Eqs.~(\ref{eq:grr}-\ref{eq:dgl}) is difficult to solve,
due to the many-body EPI. Since the EPI is weak, we consider only the lowest order Feynman diagrams
shown in Fig.~\ref{fig:fey}. The expressions for the self-energies are as follows.
The electron Fock self-energies from phonons are
\begin{eqnarray}
	\label{eq:eselg}
	\Sigma_{mn}^{F,<,>}(\varepsilon) &=& i\hbar \int M^k_{mi}G^{<,>}_{0,ij}(\varepsilon_-)D^{<,>}_{0,kl}(\omega)M^l_{jn}\frac{d\omega}{2\pi},\nonumber\\
\end{eqnarray}
\begin{eqnarray}
	\Sigma_{mn}^{F,r}(\varepsilon) &=& i\hbar \int M^k_{mi}\bigl(G^{r}_{0,ij}(\varepsilon_-)D^{<}_{0,kl}(\omega)\bigr.\nonumber\\
	&+&\bigl.G^{>}_{0,ij}(\varepsilon_-)D^{r}_{0,kl}(\omega)\bigr)M^l_{jn}\frac{d\omega}{2\pi}.
\end{eqnarray}
The Hartree self-energy does not depend on energy
\begin{eqnarray}
	\label{eq:esehar}
	\Sigma_{mn}^{r} &=& -i M_{mn}^i D^r_{ij}(\omega=0)\int M_{kl}^j G^<_{lk}(\varepsilon) \frac{d\varepsilon}{2\pi}.
\end{eqnarray}
The phonon self-energies from electrons are
\begin{equation}
	\label{eq:selg}
	\Pi^{<,>}_{mn}(\omega) = -i \int \frac{d\varepsilon}{2\pi} {\rm Tr}\left[M^mG_0^{<,>}(\varepsilon)M^nG_0^{>,<}(\varepsilon_-)\right],
\end{equation}
\begin{eqnarray}
	\label{eq:ser}
	\Pi^{r,a}_{mn}(\omega) &=& -i \int \frac{d\varepsilon}{2\pi} {\rm Tr}\left[M^mG_0^{r,a}(\varepsilon)M^nG_0^{<}(\varepsilon_-)\right.\nonumber\\
		&&+\left.M^mG_0^<(\varepsilon)M^nG_0^{a,r}(\varepsilon_-)\right],
\end{eqnarray}
with $\varepsilon_-=\varepsilon-\hbar\omega$. Summation over repeated indices is assumed here. 
Different charge and energy
conserving approximations have been developed in the literature. We will use
two of them. In Subsec.~\ref{subsec:loe}, we perform an expansion
of the current up to the second order in $M$, following the idea of
Ref.~\onlinecite{Paulsson2005}. In the numerical model calculation in Subsec.~\ref{subsec:hep}, we use
the self-consistent Born approximation (SCBA), which means we replace 
$G_0$, $D_0$ by $G$, $D$ respectively in the above equations.

\subsection{Thermoelectric transport through a single electronic level}
\label{subsec:loe}
We consider a single electronic level $H^C_{\rm e}=\varepsilon_0 d^\dagger d$,  coupled to the left and right electrodes,
characterized by the constant level-width broadening $\Gamma_\alpha$ with energy cutoff $\varepsilon_D$ (see Eq.~(\ref{eq:gm}) for the general definition). 
It interacts with an isolated phonon mode with frequency
$\omega_0$, and  $H_{\rm ep} = m_0 d^\dagger d u$.
In the linear regime, we introduce an infinitesimal change of the chemical
potential or temperature at lead $L$, e.g., $\mu_L=\mu+\delta \mu$,
$T_\sigma^L=T+\delta T_\sigma$, with $\sigma = $ e or p, $\mu$ and $T$ are the corresponding
equilibrium values.  We look at the response of the charge and heat current due
to this small perturbation. The result, up to the 2nd order in $M$, is summarized as follows,
\begin{equation}
	\left(
	\begin{array}{cc}
		\frac{I_{\rm e}}{e}\\
		I_{\rm h}\\
	\end{array}
	\right)
	=
	\left(
	\begin{array}{ccc}
		\mathcal{L}_0 & \mathcal{L}_1 \\
		\mathcal{L}_1 & \mathcal{L}_2 
	\end{array}
	\right)
	\left(
	\begin{array}{ccc}
		\delta \mu\\
		\frac{\delta T_{\rm e}}{T}
	\end{array}
	\right).
	\label{eq:lang2r}
\end{equation}
The linear conductance and the Seebeck coefficient are
\begin{eqnarray}
	G_{\rm e} &=& e^2 \mathcal{L}_0,\\
	S &\equiv& -\frac{\delta V}{\delta T} = -\frac{\mathcal{L}_1}{e\mathcal{L}_0T}. \label{eq:see}
\end{eqnarray}
The coefficients $\mathcal{L}_n$ are 
\begin{eqnarray}
	\mathcal{L}_n&=&\sum_{i=1}^3\mathcal{L}^{(i)}_n,
	\label{eq:cor}
\end{eqnarray}
with
\begin{eqnarray}
	\mathcal{L}^{(1)}_n &=& \frac{1}{\hbar}\int \frac{d\varepsilon}{2\pi} (\varepsilon-\mu)^n \left(A_{\bar{\alpha}}\Gamma_\alpha\right) f',\\
	\mathcal{L}^{(2)}_n &=& \frac{1}{\hbar}\int \frac{d\varepsilon}{2\pi} (\varepsilon-\mu)^n \left(\Delta A'_{\bar{\alpha}}\Gamma_\alpha\right)f',\\
	\label{eq:cor23}
	\mathcal{L}^{(3)}_n &=& \frac{1}{\hbar}\int \frac{d\varepsilon}{2\pi}(\varepsilon-\mu)^n \left(\Delta A''_{\bar{\alpha}}\Gamma_\alpha
	+2G_0^r{\rm Im}\Sigma_{\rm ep}^rG_0^a\Gamma_\alpha\right)f'.\nonumber\\ 
\end{eqnarray}
We have defined
\begin{eqnarray}
	\label{eq:fp}
	 f' &=& -\frac{\partial f}{\partial \varepsilon},\\
	\label{eq:gm}\Gamma_\alpha &=& i \left(\Sigma_\alpha^r - \Sigma_\alpha^a\right),\\
	\label{eq:ea}A_\alpha&=&G_0^r\Gamma^e_\alpha G_0^a,\\
	\Delta A'_{\bar{\alpha}} &=& G^r_0{\rm Re}\Sigma_{\rm ep}^rA_{\bar{\alpha}}+A_{\bar{\alpha}}{\rm Re}\Sigma_{\rm ep}^aG_0^a,\\
	\Delta A''_{\bar{\alpha}} &=& i G^r_0{\rm Im}\Sigma_{\rm ep}^rA_{\bar{\alpha}}+i A_{\bar{\alpha}}{\rm Im}\Sigma_{\rm ep}^aG_0^a,
\end{eqnarray}
and $\bar{\alpha}$ means the lead different from $\alpha$.
$\mathcal{L}^{(1)}_n$ is the single electron Landauer result.
$\mathcal{L}^{(2)}_n$ is the quasi-elastic term. $\mathcal{L}^{(3)}$ is the inelastic term. 
$f$ is the Fermi-Dirac distribution function
\begin{eqnarray}
	\label{eq:fd}f_\alpha(\varepsilon)&=&\left[{\rm exp}\left({\frac{\varepsilon-\mu_\alpha}{k_BT^\alpha}}\right)+1\right]^{-1}.
\end{eqnarray}
Since we are looking at the linear response regime, $f_L=f_R$, we dropped the subscript in Eq.~(\ref{eq:fp}).
We will also use the Bose-Einstein distribution later
\begin{eqnarray}
	\label{eq:be}n_B(\omega,T)&=&\left[{\rm exp}\left(\frac{\hbar\omega}{k_BT}\right)-1\right]^{-1}.
\end{eqnarray}
When there is no ambiguity, we will also drop the argument $T$.

In the following, we set the position of the electronic level to $\varepsilon_0 = 0$,
and look at the dependence of the conductance on the
chemical potential $\mu$. 
We firstly write down the expressions for the self-energies, and
make some observations based on their functional forms.

\begin{figure}[!htbp]
\begin{center}	
\includegraphics[scale=0.47]{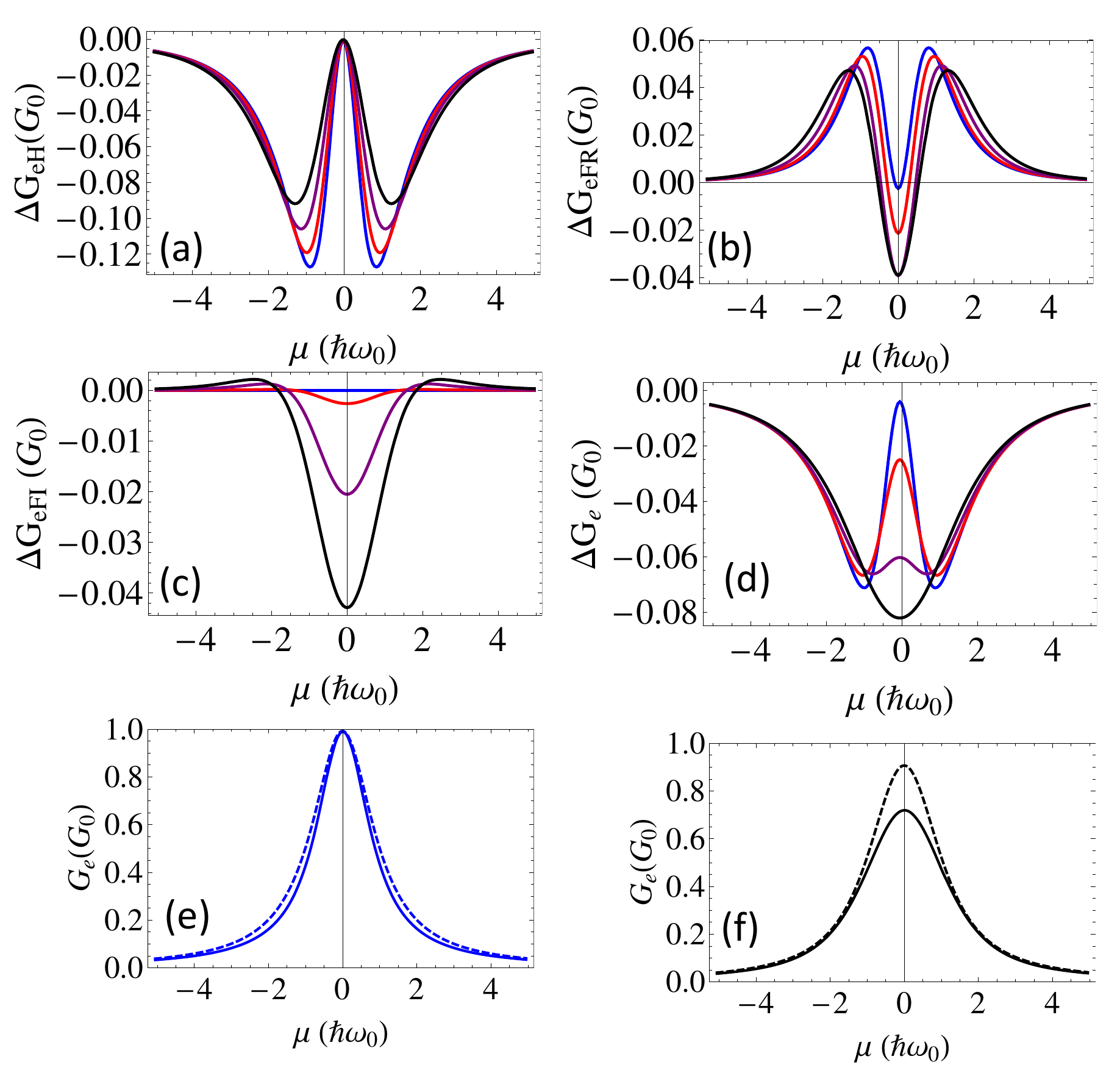}
\end{center}	
\caption{Chemical potential dependence of the conductance correction due to the
	$\Sigma^r_H (a)$, ${\rm Re}\Sigma^r_F (b)$, ${\rm Im}\Sigma^r_F (c)$,
and the sum of them ($d$). The blue, red, purple, and black lines correspond to
$k_BT=0.05, 0.15, 0.25, 0.35\hbar\omega_0$, respectively.  (e)-(f) The 
electrical conductance as a function of chemical potential ($\mu$) at $k_BT=0.05 \hbar\omega_0$ (e)
and $k_BT=0.35\hbar\omega_0$ (f). Solid lines include EPI, while the dashed lines do not. Other parameters used:
$\Gamma_L=\Gamma_R=\hbar\omega_0=1$, $m_0=1 \hbar\omega_0$/(\AA $\sqrt{\rm{u}})$.} 
\label{fig:dG}
\end{figure}

\begin{figure}[tb]
\includegraphics[scale=0.6]{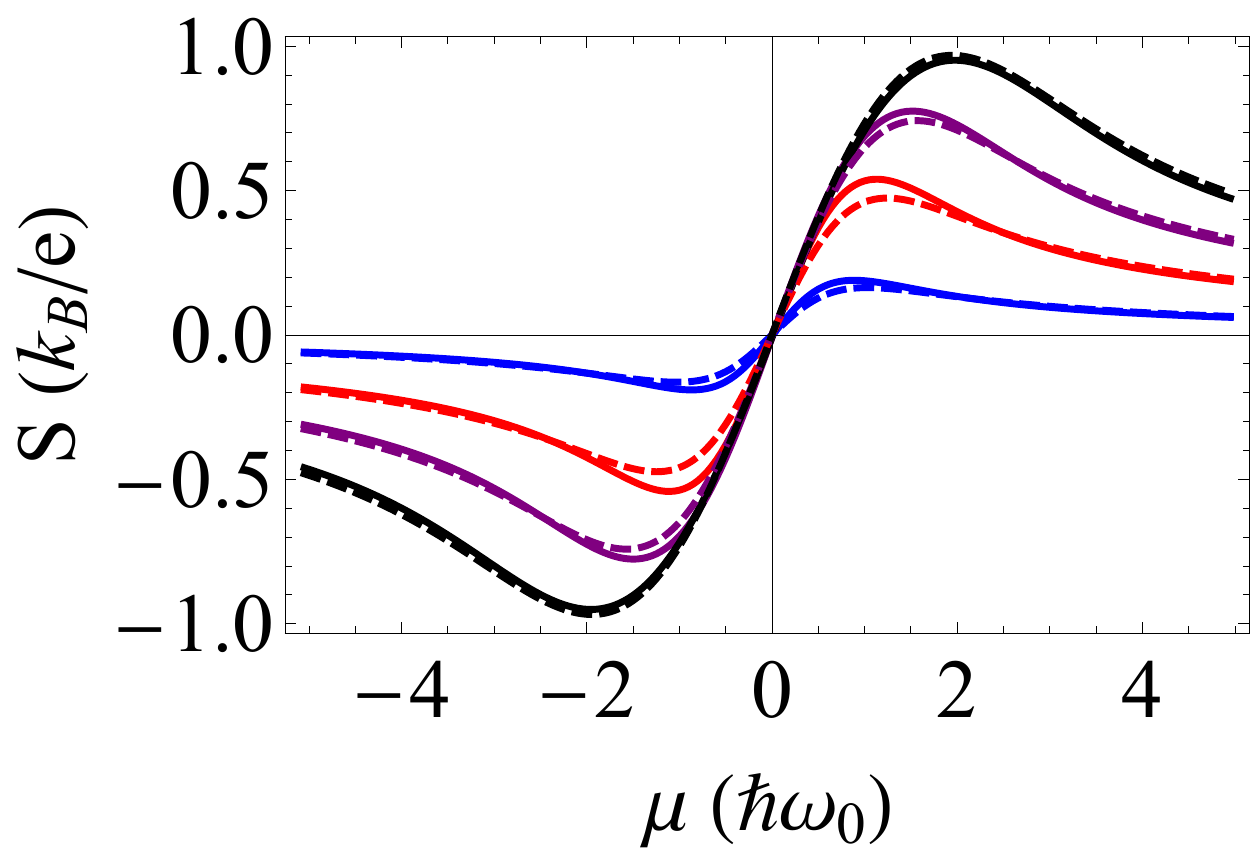}
\caption{Change of the Seebeck coefficient due to EPI at different temperatures.  The parameters and meaning of colors are the
same as Fig.~\ref{fig:dG}.}
\label{fig:sb}
\end{figure}

The Hartree self-energy is real and does not depend on
energy\cite{Lu07}
\begin{eqnarray}
	\Sigma^{r}_H &=& -\sum_\alpha \frac{m_0^2\Gamma_\alpha }{2\pi\omega_0^2}\int_{-\varepsilon_D}^{\varepsilon_D} \frac{f_\alpha(\varepsilon)}{\varepsilon^2+\Gamma^2/4}d\varepsilon.
	\label{}
\end{eqnarray}
At $T = 0$,  we get
\begin{eqnarray}
	\Sigma^{r}_H&=& -\sum_\alpha\frac{m_0^2\Gamma_\alpha}{\pi\omega_0^2\Gamma} \left. {\rm tan}^{-1}\frac{2\varepsilon}{\Gamma}\right|_{-\varepsilon_D}^{\mu_\alpha},
	\label{}
\end{eqnarray}
with  $\Gamma=\Gamma_L+\Gamma_R$.
For large
enough $\varepsilon_D$, the lower limit term turns to $-m_0^2/(2\omega_0^2)$, which
is the polaron energy shift. Note here that the $1/2$ is due to the fact we use $u_k^C$ in our definition of $H_{\rm ep}$ (Eq.~\ref{pheno-Hep}). This is different from the common definition that uses the creation and annihilation operatore $a^\dagger+a$ (Eq.~\ref{eq:hbepi}). We have subtracted the polaron shift term in the following calculation, since it is a constant. After
this subtraction, $\Sigma^r_H$ is odd in $\mu$ with a negative slope
near  $\mu=0$. We focus on the $\mu > 0$ regime. It saturates to $-m_0^2/(2\omega_0^2)$ for large $\mu$, e.g., $\mu \gg \Gamma$. At non-zero temperature, the slope and the saturation value change, but the shape of the curve is
similar to the $T = 0$ case. This means that, the Hartree
term shifts the electronic level, and reduces the conductance.
On the other hand, when $\mu \gg \Gamma$, the conductance
tends to zero, whether we include the EPI
or not. Thus, the correction to the conductance
due to the Hartree term $\Delta G_{eH} \le 0$. It starts from zero at $\mu=0$, 
goes back to zero at $\mu\gg\Gamma$, and it reaches a
maximum magnitude at some point in the middle. The described
behaviour is schematically shown in Fig.~\ref{fig:dG} (a).
This effectively reduces the broadening of the single level spectral function,
also the conductance peak.
Since the Seebeck coefficient is related to the logarithmic
derivative of the conductance, we expect it to increase
the magnitude of the Seebeck coefficient near resonance,
and to reduce it off resonance [Fig.~\ref{fig:sb}].

The imaginary part of the retarded Fock self-energy is\cite{Lu07}
\begin{eqnarray}
{\rm Im}\Sigma_F^{r}(\varepsilon) &=& -\frac{m_0^2}{4\omega_0}\sum_{\alpha,s=\pm} s A_\alpha(\varepsilon-s\hbar\omega_0)\nonumber\\
&&\times\bigl[1+n_B(s\omega_0)-f_\alpha(\varepsilon-s\hbar\omega_0)\bigr].
	\label{}
\end{eqnarray}
It is negative and even in $\varepsilon$. Its role on the differential
conductance at the phonon threshold ($eV=\hbar\omega_0$) has been discussed
extensively\cite{paulsson_unified_2008,tal_electron-vibration_2008,Egger2008,Entin-wohlman2009}.  The main conclusions are: it reduces the differential conductance at
$eV=\hbar\omega_0$ for resonant case ($\mu\sim 0$), where the bare transmission
without EPI ($T_0\sim 1$), while it does the opposite
for far off resonance case ($\mu \gg \Gamma$), where $T_0\to 0$. The transition
point between the two opposite behaviors is $T_0 = 1/2$ if the electronic
density of state (DOS) is flat. But, in general, it depends on the system
parameters. At non-zero temperature, the sharp threshold broadens out, and the
linear conductance is affected: its correction to the conductance $\Delta
G_{eFI}$ is negative for $\mu \sim 0$, and positive for $\mu\gg\Gamma$
[Fig.~\ref{fig:dG} (c)]. 

Physically, ${\rm Im}\Sigma_F^r$ gives rise to phonon scattering processes. Its
effect can be understood as follows: At $T=0$, for small bias
($eV<\hbar\omega_0$), phonon emission is not possible due to Pauli blocking,
while phonon adsorption is not possible due to zero phonon population.  So,
${\rm Im}\Sigma_F^r$ does not affect the linear conductance. At high enough
temperature, both phonon emission and adsorption are possible even at small
bias, due to the broadening of the Fermi distribution, and finite population of
phonon modes.  The phonon scattering process decreases the conductance on
resonance, but increases it far off resonance.  As a result, the Seebeck coefficient becomes smaller.

The real part ${\rm Re}\Sigma^r_F(\varepsilon)$ is obtained by the Hilbert
transform of the imaginary part. At zero temperature, it diverges
logarithmically at $\varepsilon-\mu=\pm\hbar\omega_0$\cite{Entin-wohlman2009}. Its effects on
the conductance and Seebeck coefficient are difficult to analyze. We
rely on the numerical result [Fig.~\ref{fig:dG} (b)].

Figure~\ref{fig:dG} shows the correction to the linear conductance
of different self-energy terms as a function of $\mu$. 
These numerical results confirm our qualitative
analysis. By comparing the total conductance at low [Fig.~\ref{fig:dG} (e)]
and high temperature [Fig.~\ref{fig:dG} (f)], we see that,
(1) the Hartree term dominates at low temperature, and the $G_e$-$\mu$ peak becomes
narrower. (2) the Fock term becomes important at high temperature, and the  $G_e$-$\mu$ peak
broadens out. Their effects on the 
Seebeck coefficient ($S$)  are shown in Fig.~\ref{fig:sb}. At
low temperature, when the EPI is included, the magnitude of $S$ gets larger
for $\mu \sim 0$, and smaller for $|\mu|\gg\Gamma$. At high temperature,
the effect of the Fock term results in drop of $S$. In any case, the correction to $S$
is small for weak EPI. But for the case of strong EPI,
the correction could be large (see Subsec.~\ref{subsec:qmethermoelectric}).

\subsection{Heat transport between electrons and phonons}
\label{subsec:hep}
\begin{figure}
\includegraphics[width=1.0\linewidth]{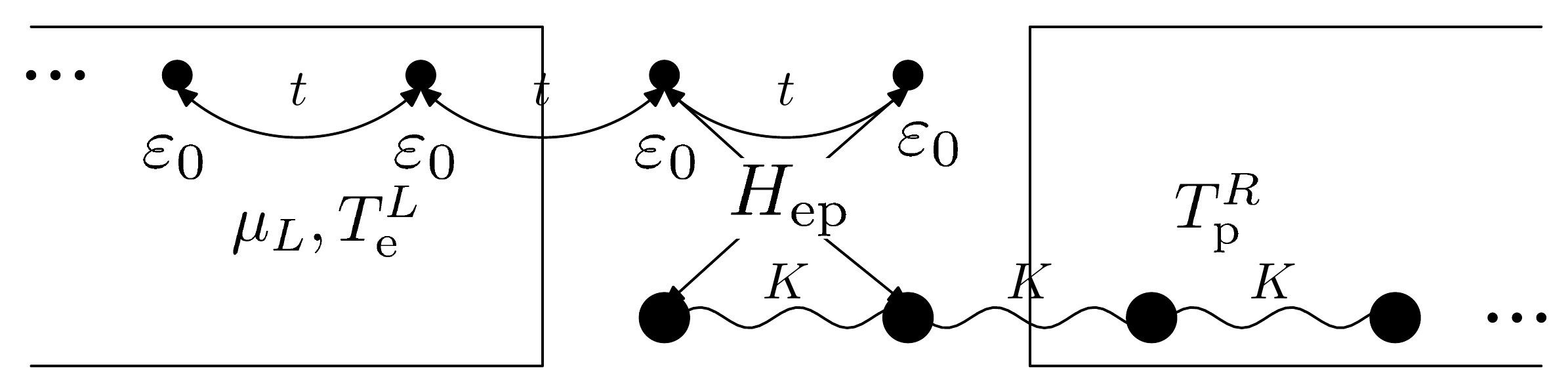}
\caption{\label{fig:mint}The model system we consider to study the energy transport between electrons and phonons.
}
\end{figure}

Let us look back at the setup in Fig.~\ref{fig:sys}. We want to study the
heat transport between electrons and phonons at finite temperature bias, but zero voltage bias. 
The simplest setup is that the system couples to one electron and one phonon
lead, each at its own temperature, see Fig.~\ref{fig:mint}.
The expression for the energy current from electrons to phonons can be obtained
from Eq.~(\ref{eq:iph}) and the expressions of the self-energies Eqs.~(\ref{eq:eselg}-\ref{eq:ser})\cite{Lu07}
\begin{eqnarray}
	\label{eq:heatgen}
	Q&=&i\hbar\int \frac{d\varepsilon}{2\pi} \int \frac{d\omega}{2\pi} \omega
	\left[G^>_{nm}(\varepsilon)M_{mi}^kD_{kl}^<(\omega)G^<_{ij}(\varepsilon_-)M_{jn}^l\right],\nonumber\\
\end{eqnarray}
where, again, summation over repeated indices is assumed.
For the ease of analysis, we perform an expansion of the
above expression to 2nd order in $M$, and it becomes
\begin{eqnarray}
	Q^{(2)} &=&\hbar \int_{0}^{+\infty}\frac{d\omega}{2\pi} \omega {\rm Tr}\bigl[ \Lambda(\omega,T_{\rm e})\mathcal{A}(\omega) \bigr]\nonumber\\
	&&\times \bigl[ n_B(\omega,T_{\rm e})-n_B(\omega,T_{\rm p}) \bigr],
	\label{eq:ephp}
\end{eqnarray}
where
\begin{eqnarray}
	\Lambda(\omega,T_{\rm e}) &=& \int \frac{d\varepsilon}{2\pi}{\rm Tr}\bigl[ MA(\varepsilon)MA(\varepsilon_-) \bigr]\nonumber\\
	&\times&\bigl[f(\varepsilon,T_{\rm e})-f(\varepsilon_-,T_{\rm e})\bigr],
	\label{}
\end{eqnarray}
and
\begin{eqnarray}
	\mathcal{A}(\omega)&=&i\bigl[D_0^r(\varepsilon) - D_0^a(\varepsilon)\bigr].
	\label{eq:sf0}
\end{eqnarray}
Now the question we ask is whether there is a diode behaviour for the heat
transport between electrons and phonons\cite{li_colloquium:_2012}, e.g., $Q(\Delta T)\neq Q(-\Delta T)$, with $\Delta T=T_{\rm e}-T_{\rm p}$. 
This is relevant because in some
special situation, e.g., at metal-insulator interface, or insulating molecular
junctions, EPI becomes the bottleneck of heat
transfer\cite{majumdar_role_2004,segal_single_2008,renjie13,lifa13,Vinkler14}.  We can define the
rectification ratio as
\begin{equation}
	R = \frac{Q(\Delta T)+Q(-\Delta T)}{Q(\Delta T)-Q(-\Delta T)}.
	\label{eq:dr}
\end{equation}
If we assume a constant electron DOS, $\Lambda(\omega)$ does not
depend on $T$, we get $Q(-\Delta T)=-Q(\Delta T)$, and $R=0$. The physical
reason is that in this case, it is possible to map the
electron-hole pair excitation into harmonic
oscillators\cite{Tomonaga50,Luttinger63,segal_single_2008}. Then, it is equivalent to
heat transport within a two-terminal harmonic system. We do not expect any
rectification effect. To make $R\neq0$, the electronic DOS has to be
energy-dependent within the broadening of the Fermi-Dirac distribution given by
$k_B\Delta T$. This effectively introduces anharmonicity into the system,
consistent with previous studies\cite{segal_single_2008,renjie13,lifa13}.

We can go one step further, by making a Taylor expansion
of the spectral function $A(\varepsilon)$ about the Fermi
energy, we find that the sign of $R$ is determined by the
sign of $\frac{\partial^2 A}{\partial \varepsilon^2}$ (The 1st
order term is zero).

\begin{figure}[tb]
\includegraphics[scale=0.5]{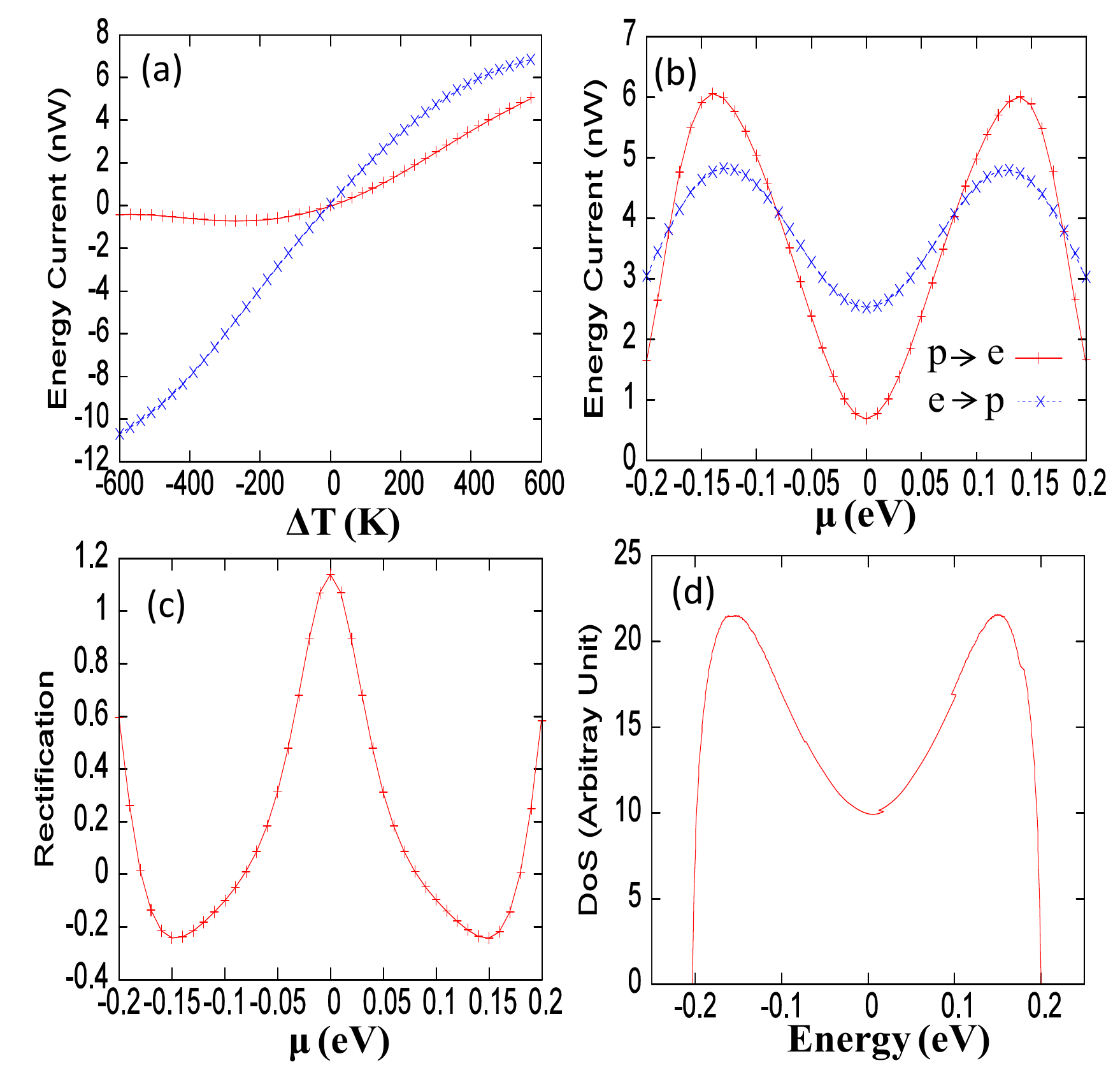}
\caption{(a) The energy current with $\mu = 0$ (red, solid) and $\mu = 0.1$ eV (blue, dotted).
	We have set $T_{\rm e} = T+\Delta T/2$, $T_{\rm p} = T-\Delta T/2$, with $T = 300$ K.  (b) Fermi level ($\mu$) dependence of the energy current at fixed temperature difference
$\Delta T = 300$ K.  (c) Rectification ratio $R$ as a function of $\mu$.  (d)
The electronic spectral function $A(\varepsilon)$ as a function of energy
$\varepsilon$.} 
\label{fig:rect}
\end{figure}

To check this argument, we calculated the heat current
across one-dimensional (1D) metal-insulator junction.
The metal side is represented by a 1D tight-binding
chain, with hopping element $t=-0.1$ eV, and onsite
energy $\varepsilon_0=0$. The insulator side is represented by a
1D harmonic chain with the spring constant $K = 0.1$ eV/(\AA$^2$u). 
The insulator and metal couple through
their last two degrees of freedom. Their interaction matrices
are
\begin{equation}
	M^k = (-1)^k m_0 \left(
	\begin{array}{cc}
		0 & 1 \\
		1 & 0 
	\end{array}
	\right), \quad k = 1,2.
\end{equation}
Here, $m_0=0.05$ eV/(\AA $\sqrt{\rm{u}}$), $k$ represents the phononic degrees
of freedom.  This means that the system couples to only one electron and one
phonon lead [Fig.~\ref{fig:mint}].  Figure~\ref{fig:rect} summarizes our
result. In Fig.~\ref{fig:rect} (a), we show the energy current as a function of
temperature difference between the metal and insulator ($\Delta T$), at two
Fermi levels ($\mu = 0, 0.1$ eV). The energy current is asymmetric with respect
to the sign change of $\Delta T$. But the rectification ratio $R$ has opposite
sign. This is further highlighted in the plot of the energy current and the
rectification ratio $R$ as a function of $\mu$ for fixed temperature bias
$\Delta T = \pm 300$ K [Fig.~\ref{fig:rect} (b) and (c), respectively]. The
sign of $R$ is well correlated with the sign of $\frac{\partial^2 A}{\partial
\varepsilon^2}$ [Fig.~\ref{fig:rect} (d)].

\subsection{Effect of EPI on thermal transport in single-walled carbon nanotubes}
Phonon modes can be excited by the mobile electrons due to the EPI effect;
i.e., a high bias over the system leads to self-heating (Joule heating). In nanoscale electric devices, the electric current density can be much larger than
that in the macroscopic system. The high current density will
generate strong Joule heating, which may eventually break the device. In this
sense, Joule heating becomes a bottleneck for further increase of the
electric current density. Hence, lots of theoretical and experimental efforts
have been devoted to understanding the Joule heating phenomenon in the
nanoscale electric devices. In experiment, Joule heat can be measured via the
thermal-mechanical expansion technique, which records the Joule heat induced
temperature rise\cite{GrosseKL2013apl}. The Joule heat can increase the
temperature of the molecular junction from room temperature to 463~K, which
has been examined through the inelastic electron tunneling
spectroscopy\cite{TsutsuiM2008nl}. Grosse et al. investigated the nanoscale
Joule heating in phase change memory devices\cite{GrosseKL2013apl}. The Joule
heating leads to the temperature rise in the phase change memory device,
which results in an obvious volume expansion.  In another experiment, 
Joule heating is found to be responsible for the correlated breakdown of
nanotube forests\cite{ShekharS2011apl,ShekharS2012carbon}.

For a system without localized phonon modes, all phonon modes have important
contribution to the Joule heat.  The Joule heat contributed by these propagating phonon
modes have important effects on the electric devices. For example, in
graphene transistors, the output electric current will saturate with
increasing source-drain voltage\cite{LiaoAD2011prl,IslamS2013ieee}. This
saturated current density can be reduced by 16.5\% due to the Joule
heating\cite{IslamS2013ieee}.

The localized phonon modes exist around some defects or nonuniform
configurations, such as the free edge, the isotropic doping, interface, 
etc. This particular type of phonon modes has no direct contribution to the
thermal conduction, but localized phonon
modes play a particularly important role in the Joule heat phenomenon. They
are characteristic for their exponentially decaying vibration displacement; i.e., only a small portion of atoms are involved in the localized vibration. For instance, there are some localized edge phonon modes
at graphene nanoribbon's free edge. In these modes, edge atoms vibrate with
large amplitude, but the vibrational displacement decays exponentially from the edge into the interior region.

The localized-phonon-mode-induced Joule heat was observed in 
graphene nanoribbons in experiment, and explained theoretically. Jia, et al.
utilized Joule heating to trigger the edge reconstruction at the free edges
in the graphene nanoribbons\cite{JiaX2009sci}. Engelund, et al. attributed this phenomenon to the Joule heating of the edge phonon
modes\cite{EngelundM2010prl}. There are two conditions for the important Joule
heating of the edge phonon modes. First, these localized edge phonon modes can
spatially confine the energy at the edges. Second, the electrons interact strongly with the localized edge phonon modes. The mean steady-state occupation of the
edge phonon mode can be calculated from the ratio of the current-induced phonon
emission rate and damping rate. The effective temperature for the free edge can
be extracted by assuming this occupation to be Bose distributed. The effective
temperature was found to be as high as 2500~K for bias around 0.55~V. This high
effective temperature was proposed to be the origin for the edge
reconstruction.

Although Joule heating might be used for selectively bond-breaking\cite{KochJ2006prb,VolkovichR2011pccp},
its most common outcome is a
disaster of device breakdown. The effective temperature is a suitable quantity to
describe the Joule heating. In 1998, Todorov studied Joule heating problem in a
molecular junction\cite{To.1998}. In his work, the Einstein model is
applied to represent the phonon modes in the system, and the electron-electron
interaction is ignored as the system size is much smaller than the electron
mean free path. Part of the EPI-induced Joule heat will be delivered out of the
system by the phonon heat conduction, while the remaining Joule heat gives a high effective temperature. For low ambient temperature, the effective temperature
scales with voltage $V$ as $T_{\rm eff}^{4}\approx \gamma^4 V^2$, with $\gamma$
as an EPI-dependent constant. It was shown that the effective temperature can
be above 200~K for a very low ambient temperature around
4~K\cite{AsaiY2011prb}. But at very high bias, the scaling law could differ from this\cite{TsutsuiM2007apl}.

There are several experimental approaches to investigate the effective
temperature of the electric device induced by Joule heating. The effective
temperature can be extracted by measuring some quantities that are
temperature-dependent. For example, the breaking force of the single
molecular junction is related to the temperature. This force-temperature
relationship can be used to estimate the effective
temperature\cite{HuangZ2006nl}. The Raman spectroscopy also depends on the
temperature. Hence, it can be used to deduce the effective temperature of
Raman-active phonon modes\cite{IoffeZ2008nn,YuYJ2011apl,WardDR2011nn}.

\begin{figure}[tb]
  \begin{center}
    \scalebox{1.0}[1.0]{\includegraphics[width=7cm]{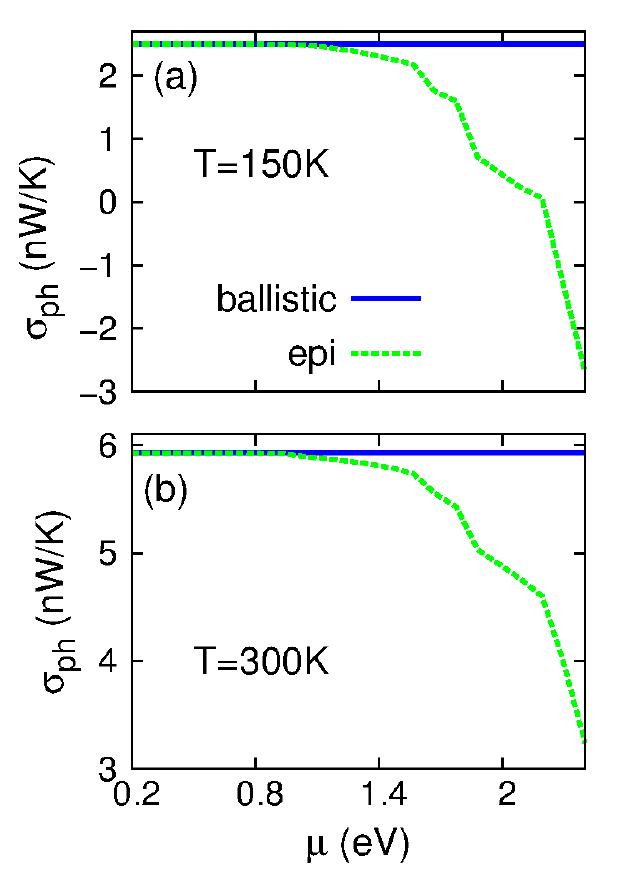}}
  \end{center}
  \caption{(Color online) The phonon thermal conductance versus chemical potential for metallic SWCNT (10, 10) at (a) 150~K and (b) 300~K. Solid line is for ballistic phonon thermal conductance without EPI effect. Reprinted from J. Appl. Phys., {\bf 110}, 124319 (2011).}
  \label{fig_t_a_sgm_ph_u}
\end{figure}

\begin{figure}[tb]
  \begin{center}
    \scalebox{1.0}[1.0]{\includegraphics[width=7cm]{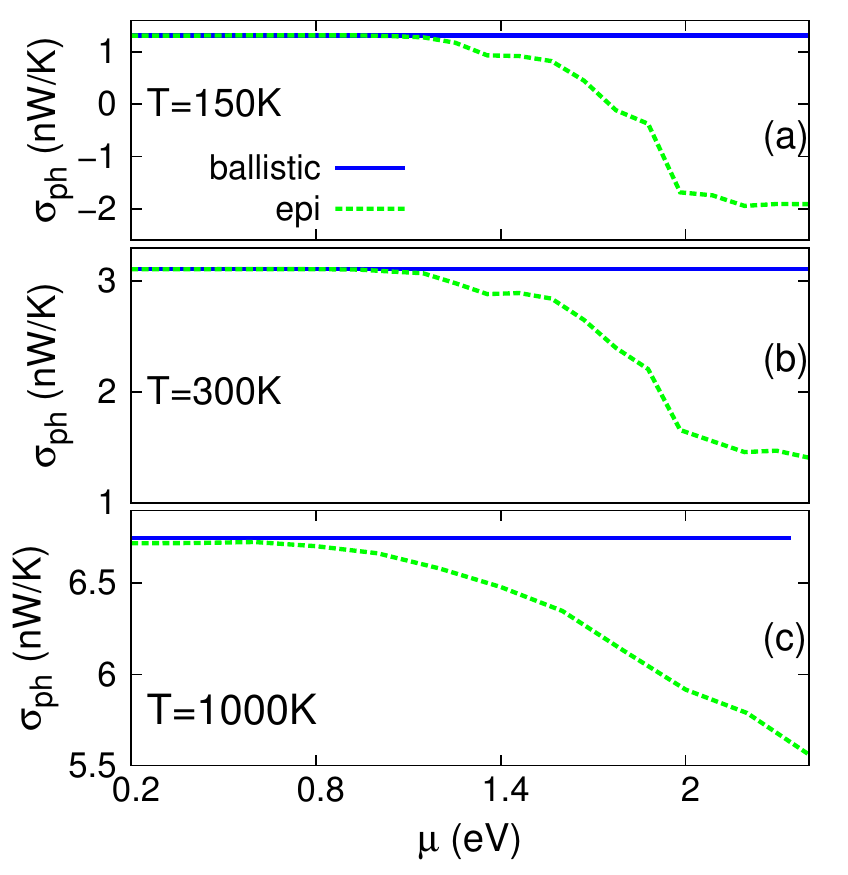}}
  \end{center}
  \caption{(Color online) The phonon thermal conductance versus chemical potential for semiconductor SWCNT (10, 0) at (a) 150~K, (b) 300~K, and (c) 1000~K. Solid line is for ballistic phonon thermal conductance without EPI effect.  Reprinted from J. Appl. Phys., {\bf 110}, 124319 (2011).}
  \label{fig_t_z_sgm_ph_u}
\end{figure}

It has also been shown that EPI has an important effect on the thermal
conductance in single-walled carbon nanotubes (SWCNTs)\cite{JiangJW2011joule}. 
For them, we apply the Born approximation to consider the EPI effect using the
NEGF approach, as the SCBA is computationally more expensive. The phonon
thermal current can be calculated by considering the three EPI contributions
shown in the Feynman diagrams in Fig.~\ref{fig:fey}. The phonon thermal
current flowing from the left lead into the center is given by
Eq.~(\ref{eq:iph}). The expression for the right lead is analogous. The Joule heat is generated
in the system and flows into the leads, so the total Joule heat is the sum of
heat currents into both leads, $Q=-(I_{\rm p}^{L}+I_{\rm p}^{R})$. The thermal current from Eq.~(\ref{eq:iph}) also includes that induced by the temperature
gradient, which satisfies $I_{\rm p}^{L}=-I_{\rm p}^{R}$. Hence, $Q$ gives
solely the Joule heat.For metallic SWCNT (10, 10),
both electrons and phonons are important heat carriers. The EPI only slightly
reduces the electron thermal conductance, but it has a strong effect on the
phonon thermal conductance. More specifically,
Fig.~\ref{fig_t_a_sgm_ph_u}~(a) shows an `electron-drag' effect on the phonon
thermal conductance at 150~K. The phonon thermal conductance becomes negative
for high chemical potential value $\mu > 2.0$~{eV}, which indicates that
electrons can help to drag phonons from cold temperature region to the hot
temperature region. The `electron-drag' phenomenon happens at low temperature
and high chemical potential, and it does not happen at a higher temperature
300~K as shown in Fig.~\ref{fig_t_a_sgm_ph_u}~(b). For semiconductor SWCNT
(10, 0), the electronic thermal conductance contributes less than 10\% of the
total thermal conductance at low bias (e.g. $\mu=0.3$~{eV}), while phonons
make most significant contribution to the total thermal conductance. Similar
`electron-drag' phenomenon also exists in the semiconductor SWCNT (10, 0) at
low temperature as shown in Fig.~\ref{fig_t_z_sgm_ph_u}~(a).
\section{Strong EPI regime: Quantum Master equation approach} \label{sec:qme}
\subsection{Quantum Master equation formulism} In this section, we introduce
the QME approach to consider the case of strong EPI. Before doing that, we
should mention that the NEGF method has also been used to treat the strong
EPI\cite{Flensberg2003,Galperin2004,Chen2005,Galperin2006,Sun2007,Zazunov2007}. Since the idea behind it is very similar to that of the master equation
approach, we choose not to introduce it here. 

To simplify the formula, we ignore the coupling of molecular phonon modes to
the phonon leads.  The model Hamiltonian simplifies to
\begin{equation}
	H_{\rm tot}=\Hs+H_{\rme}^L+H_{\rme}^R+{\Hse}
\end{equation}
where $\Hs=H_{\rmpp}^C+H_{\rme}^C+H_{\rmep}$ denotes the system Hamiltonian,  and $\Hse=V_{\rme}^{L}+V_{\rme}^{R}$ is the system-lead coupling. In the QME formalism we
assume the system-lead coupling  $\Hse$ is weak so we can do perturbation on
it. We work in the interaction picture with $\Ho=H-\Hse$ as
non-interacting part and $\Hse$ as the interaction. For simplicity, in this section we use $V$ to represent $\Hse$ since we don't have $V_{\rm p}$. The equation of motion for the
full density matrix follows the von Neumann equation
\begin{equation}
i\hbar\frac{\partial\irho(t)}{\partial t}=[\iV(t),\irho(t)].
\end{equation}
Here, the subscript $I$ denotes operator in the interaction picture.
The time argument in the parentheses means non-interacting evolution $O(t)=\Ud{\Ho}{t}O\U{\Ho}{t}$. The above equation can be written in an integral form as
\begin{equation}
\label{eq:integration form}
\irho(t)=\frac{-i}{\hbar}\int_{t_0}^t\,d\tauu [\iV(\tauu),\irho(\tauu)]+\irho(t_0).
\end{equation}
One can recursively apply the above equation to get a series expansion of the full density matrix in power of $\iV$. We truncate the series to the second order and differentiate it with respect to time $t$ at both sides of the equation to get the following integro-differential equation

\begin{equation}
	\frac{\partial\irho(t)}{\partial t}{\approx}\frac{-i}{\hbar}[\iV(t),\irho(t_0)]-\frac{1}{\hbar^2}\int_{t_0}^td\tauu[\iV(t),[\iV(\tauu),\irho(t_0)]].
\end{equation}
We prepare the initial state as a product state of the system and each lead, $\frho(t_0)=\rrho(t_0)\otimes\rho_{\rm e}^{L}\otimes\rho_{\rm e}^{R}$. For the system-lead coupling $V$ we assume it can be written as a product of system operator $S$ and lead operator $B$ as $V=\sum_{\alpha}S^\alpha\otimes\E^\alpha$. In such cases we can trace over the lead degrees of freedom to get

\begin{equation}
\frac{\partial\ir(t)}{\partial t}=\frac{-1}{\hbar^2}\sum_{\alpha,\beta}\int_{t_0}^t\!\!d\tauu [\iS^\alpha(t), \iS^\beta(\tauu)\ir(t_0)]C^{\alpha\beta}(t-\tauu)+\hc
\end{equation}
where $C^{\alpha\beta}(t-\tauu)=\mathrm{Tr}[\rho_{\rm e}^{L}\otimes\rho_{\rm e}^{R}\iE^\alpha(t)\iE^\beta(\tauu)]$ is the correlation function of the leads. Here we have used the condition that the expectation value of a single lead operator $B^\alpha$ is zero. We can now transform back to the \schrodinger picture and extend the initial time $t_0$ to $-\infty$ to get the QME of Redfield type\cite{Redfield1965,breuer2007theory,weiss}
\begin{eqnarray}
\label{eq:rho}
\frac{\partial\rrho}{\partial t}&=&-\frac{{i}}{\hbar}[H_S,\rrho]-\frac{1}{\hbar^2}\sum_{\alpha ,\beta}\\
&&\times\int_{-\infty}^td\tauu\bigl\{[S^\alpha,S^\beta(\tauu-t)\rrho]C^{\alpha\beta}(t-\tauu)+\hc\bigr\}. \nonumber
\end{eqnarray}
Here we have replaced $\rrho(t_0)$ by $\rrho$, which is essential and correct
only when one intends to get the 0th-order reduced density matrix $\tilde{\rho}^0$ by solving the
above QME\cite{Thingna2014,WaAgLiTh14,Fleming11}. In the application to the EPI problem, by exact diagonalizing the system Hamiltonian, this Redfield QME can take into account the coherence between electrons and phonons, in contrast to the usual rate equation approach\cite{Segal05,Wu09}.

We write the above equation in the eigenbasis of the system Hamiltonian $H_S$ to obtain\cite{breuer2007theory,Zwanzig}
\begin{equation}
	\frac{d\rrho_{nm}}{dt}=-\frac{{i}}{\hbar}\Delta_{nm}\rrho_{nm}+\sum_{ij}R^{ij}_{nm}\rrho_{ij},
\end{equation}
where the relaxation tensor reads\cite{Juzar2012}
\begin{eqnarray}
R^{ij}_{nm}&=&\frac{1}{\hbar^2}\sum_{\alpha ,\beta}\Bigl\{ S^\alpha_{ni}S^\beta_{jm}W_{ni}^{\alpha\beta}\nonumber\\
                    &&-\delta_{jm}\sum_lS^\alpha_{nl}S^\beta_{li}W_{li}^{\alpha\beta}\Bigr\}+\mathrm{H.c.}
\end{eqnarray}
The  transition coefficients are given by
\begin{equation}
W_{kj}^{\alpha\beta}=\int_{-\infty}^{t} d\tauu e^{{i}\Delta_{kj}(\tauu-t)/\hbar}C^{\alpha\beta}(t-\tauu),
\end{equation}
where $\Delta_{kj}=E_k-E_j$ are the energy spacings of the system Hamiltonian.

Since we are only interested in the steady state, we impose the condition $d\rrho/dt=0$ at $t=0$ and solve the above equation order by order with respect to $V$. One can find that all the off-diagonal elements of the 0th-order reduced density matrix vanish in steady state and the diagonal elements can be evaluated via the matrix equation\cite{Juzar2012} $\sum_iR^{ii}_{nn}\rrho^{(0)}_{ii}=0$, together with the constraint of $\mathrm{Tr}[\rrho^{(0)}]=1$.

For the calculation of currents, we go through a similar derivation as the QME. 
The electronic current operator $\ob_{\rm e}$ and heat current
operator $\ob_{\rm h}$ can be written in the form $\ob_{\rm e(h)}=\sum_\alpha
S^\alpha\otimes\Ej_{\rm e(h)}^\alpha$. The expectation value of currents can be
calculated according to $I_{\rm e(h)}=\tr[\irho(t)\io(t)]$. Since we are interested
in the lowest order of current, we truncate Eq.~(\ref{eq:integration form}) to the
lowest order and plug in to get
\begin{equation}
I_{\rm e(h)}=\frac{-i}{\hbar}\int_{t_0}^t\,d\tauu \tr\bigl\{[\iV(\tauu),\irho(t_0)]\io(t)\bigr\}.
\end{equation}
By taking the trace over the leads and transforming back to the \schrodinger picture one obtains
the current at $t=0$ as
\begin{eqnarray}
	I_{\rm e(h)}&=&\frac{1}{\hbar^2}\sum_{\alpha,\beta}\int_{-\infty}^{0}dt' \tr[\rrho S^\alpha S^\beta(t')]\mathcal{C}_{\rm e(h)}^{\alpha\beta}(-t')+\mathrm{H.c.},\nonumber\\
\end{eqnarray}
where $\mathcal{C}_{\rm e(h)}^{\alpha\beta}(t)=\bigl\langle
B^\alpha(t)\mathcal{B}_{\rm e(h)}^\beta(0)\bigr\rangle$ is the correlation function
between the lead operators occurring in the system-lead coupling Hamiltonian and the current operator definition. For the same reason as the derivation of
master equation, here $\rrho(t_0)$ needs to be replaced by $\rrho$ to get
correct steady state results. Since we are calculating lowest order of current,
we can use the 0th order reduced density matrix $\rrho^{(0)}$. Written in the eigenbasis of system Hamiltonian, the above equation becomes
 $I_{\rm e(h)} =
\mathrm{Tr}\left[\rrho^{(0)}\mathcal{I}^r_{\rm e(h)}\right]$ with the reduced
current operator defined as
\begin{equation}
	(\mathcal{I}^r_{\rm e(h)})_{ij}=\frac{1}{\hbar^2}\sum_{\alpha,\beta, k}\Bigl[S^\alpha_{ik}S^\beta_{kj}\mathcal{W}^{\alpha\beta}_{\rm e(h)}(\Delta_{kj})+{\rm c.c.}\Bigr],
\end{equation}
where the transition coefficients are
\begin{equation}
\mathcal{W}_{\rm e(h)}^{\alpha\beta}(\Delta_{kj})=\int_{-\infty}^0 d\tau e^{i\Delta_{kj}\tau/\hbar}\mathcal{C}_{\rm e(h)}^{\alpha\beta}(-\tau).
\end{equation}.

Up to now the QME formalism is general and not restricted to any specific form
of system or leads Hamiltonians. For the application to transport problems with
EPI as concerned in this review, the system-coupling Hamiltonian is considered
as a tunneling Hamiltonian\cite{MeirWingreen}. In such case the system operator
$S$, leads operator $\E$ and $\Ej$ can be specified as
$S=\left\{d,d^\dagger\right\}$, $B=\left\{ \sum_{k\in L,R} V_k c_k^\dagger,
\sum_{k\in L,R} V_k c_k \right\}$, $\mathcal{B}_{\rm e}=\left\{\,e\sum_{k\in
L}V_kc_k^\dagger,-e\sum_{k\in L}V_kc_k\,\right\}$ and
$\mathcal{B}_{\rm h}=\left\{\,\sum_{k\in L}(\varepsilon_k-\mu_L)
V_kc_k^\dagger,-\sum_{k\in L}(\varepsilon_k-\mu_L) V_kc_k\,\right\}$. The
infinite nature of the leads can be specified by defining a continuous spectra
function for the leads
\begin{eqnarray}
	\Gamma_\alpha(\varepsilon)&=&-2{\rm Im}\Sigma_\alpha^r (\varepsilon)\nonumber\\
	&=& 2\pi\sum_{k\in\alpha}|V_k|^2\delta(\varepsilon-\varepsilon_k),\;\alpha=L,R.
\end{eqnarray}
Throughout this section we use a wide band spectra function for the electronic leads with Lorentzian cut-off as
\begin{eqnarray}
\label{eq:spece}
\Gamma_\alpha(\varepsilon)&=&\frac{\eta_\alpha}{1+(\varepsilon/\varepsilon_D)^2},\;\alpha=L,R. 
\end{eqnarray}
The non-vanishing correlation functions can be evaluated via
\begin{eqnarray}
C_\alpha^{12}(t)&=&\int_{-\infty}^\infty \frac{d\varepsilon}{2\pi}\Gamma_\alpha(\varepsilon)f_\alpha(\varepsilon)e^{\mathrm{i}\varepsilon t/\hbar},\\
C_\alpha^{21}(t)&=&\int_{-\infty}^\infty \frac{d\varepsilon}{2\pi}\Gamma_\alpha(\varepsilon)\bigl(1-f_\alpha(\varepsilon)\big)e^{-\mathrm{i}\varepsilon t/\hbar}\!\!, \\
\mathcal{C}_{\rm h}^{12}(t)&=&\int_{-\infty}^\infty \frac{d\varepsilon}{2\pi}(\varepsilon-\mu_{L})\Gamma_L(\varepsilon)f_L(\varepsilon)e^{\mathrm{i}\varepsilon t/\hbar},\\
\mathcal{C}_{\rm h}^{21}(t)&=&\int_{-\infty}^\infty \frac{d\varepsilon}{2\pi}(\varepsilon-\mu_{L})\Gamma_L(\varepsilon)\bigl(1-f_L(\varepsilon)\big)e^{-\mathrm{i}\varepsilon t/\hbar},\nonumber\\
\end{eqnarray}
and $C^{12}(t) = C^{12}_L(t)+C^{12}_R(t)$, $\mathcal{C}_{\rm e}^{12}(t)=-eC_L^{12}(t)$, $\mathcal{C}_{\rm e}^{21}(t)=eC_L^{21}(t)$. We note that here the upper index $1$ or $2$ refers to the two components of $B$ and $\mathcal{B}_{\rm e/h}$ given above.
For the system Hamiltonian, we focus only on the single electronic level coupled to a single phonon mode representing the center of mass of the molecule. In such case, the system Hamiltonian will reduce to
\begin{equation}
\Hs=\varepsilon_0d^\dagger d+\hbar\omega_0a^\dagger a+\lambda d^\dagger d(a^\dagger+a)
\label{eq:hbepi}
\end{equation}
with $\omega_0$ the angular frequency of the phonon mode and $\lambda$ denotes the EPI strength. 
This type of Hamiltonian has been well-studied in the context of molecular junction\cite{Chen2005,Koch2005,Sun2007,Lu07,McEniry2008a,Asai08,Lee2009,Leijnse2010,Zianni2010,Piovano2011,Koch2011,Zhou2012,Arrachea2014,Koch2014,Perroni2014}.

The QME formalism treats the nonlinearity of EPI exactly. Therefore in this section we will focus on strong EPI regime with emphasis on the EPI strength dependence of the electron/heat current. In the following we will discuss the effect of EPI on the electronic transport properties,  including the phonon sidebands, negative differential resistance, thermoelectric properties and local heating effects.

\subsection{Phonon sidebands and negative differential resistance}
One of the earliest findings of the vibrational effects on the electronic
transport through a molecular quantum dot is the appearance of the phonon
sidebands in the $I-V$ characteristics. When electrons transport through a
single electronic level, the differential conductance ($dI/dV$) will manifest a
peak at resonant level when plotted against the voltage bias ($V$). However,
when the electronic level is interacting with a vibrational mode, replica side
peaks will appear at the side of the resonant peaks. These side peaks are called
phonon sidebands. A simple reason for the appearance of phonon sidebands is due to the fact that the electrons can emit or absorb phonons when they pass through the molecule. Therefore, the distance between each adjacent peaks is always equal to a single phonon energy. The phonon sidebands attract wide attentions in molecular
junction systems. Experimentally, phonon sidebands were found in
1980s\cite{Goldman1987} and then were utilized to identify vibrational modes
in molecular junctions \cite{Park2000,Park2002,Zhitenev2002,Yu2004} and quantum wires\cite{Agrait2002,Agrait2002b,DeLaVega2005,Paulsson2005,ViCuPaHa05}.  Theoretically, at the beginning the sidebands
were investigated by using scattering theory, which gives the transmission
probability $T({\varepsilon,\varepsilon'})$ for an electron to passing through an EPI system. The electron is coming from vacuum at energy ${\varepsilon}$ and leaving at energy ${\varepsilon'}$\cite{Wingreen1988}. The scattering theory predicts side peaks in the
transmission probability, which qualitatively justifies the phonon sidebands in
molecular junction systems. However, prediction of phonon sidebands in the
lead-molecule-lead junctions is a much more difficult task. A simple
generalization to take the Fermi-Dirac statistics nature of the electron leads
into account is to weight the exact transmission probability with the
Fermi-Dirac distribution of each lead, i.e., by multiplying the transmission
probability $T({\varepsilon,\varepsilon'})$ by
a factor $f_L({\varepsilon})[1-f_R({\varepsilon'})]$ as a new transmission
probability for electron going from the left lead to the right lead through the nano-conductor. 
This approximation is called single
particle approximation (SPA). Plenty of earlier work is in this framework\cite{} and
it predicts Lorentzian type of phonon sidebands with the same width. But
obviously this method assumes each electron transports independently through
the junction, where the many-body effects are ignored. As a result, it
overestimates the currents and it is not able to predict the quantized
conductance $e^2/h$ either.

Based on the NEGF technique, more rigorous methods merged in dealing with the nonlinearity in EPI, such as the Green's function equation of motion method (EOM) \cite{Flensberg2003}, SCBA\cite{MiAlMi04,Chen2005,Galperin2006,Galperin2004} and nearest neighbor crossing approximation (NNCA)\cite{Zazunov2007}. All of the above are Green's function based formalisms with different kinds of approximations. In general these approaches predict that the phonon side peaks are much sharper than the SPA approach. This sharpness is closely related to the Pauli exclusion, which the SPA approach failed to take into account. Other than Green's function based methods, another approach is to use rate equation of electron occupation probability in the molecule, via calculating the transition coefficients of the electron to tunnel from the molecule to each lead and vice versa\cite{}. This method assumes the transport is an electron tunneling process and the electron will lose its phase information when it resides in the molecule. Therefore it will be valid when the molecule-lead coupling is weak and the coherence of the electron and phonon in the molecule can be neglected. For all these formalisms, we would like to point out that one should take care of the phonon distribution. Treating the phonon at equilibrium distribution\cite{} at a fixed temperature could be valid when the EPI strength is much weaker than the coupling strength between the phonon and its environment. However, when the environmental influence is weak, one should consider phonons in nonequilibrium states. This nonequilibrium treatment of phonon distribution can have pronounced effect on $I-V$ characteristics due to the fact that the current induced vibrational excitation can be significant\cite{MiAlMi04, Hartle2011b}.

\begin{figure}
\includegraphics[width=1.0\linewidth]{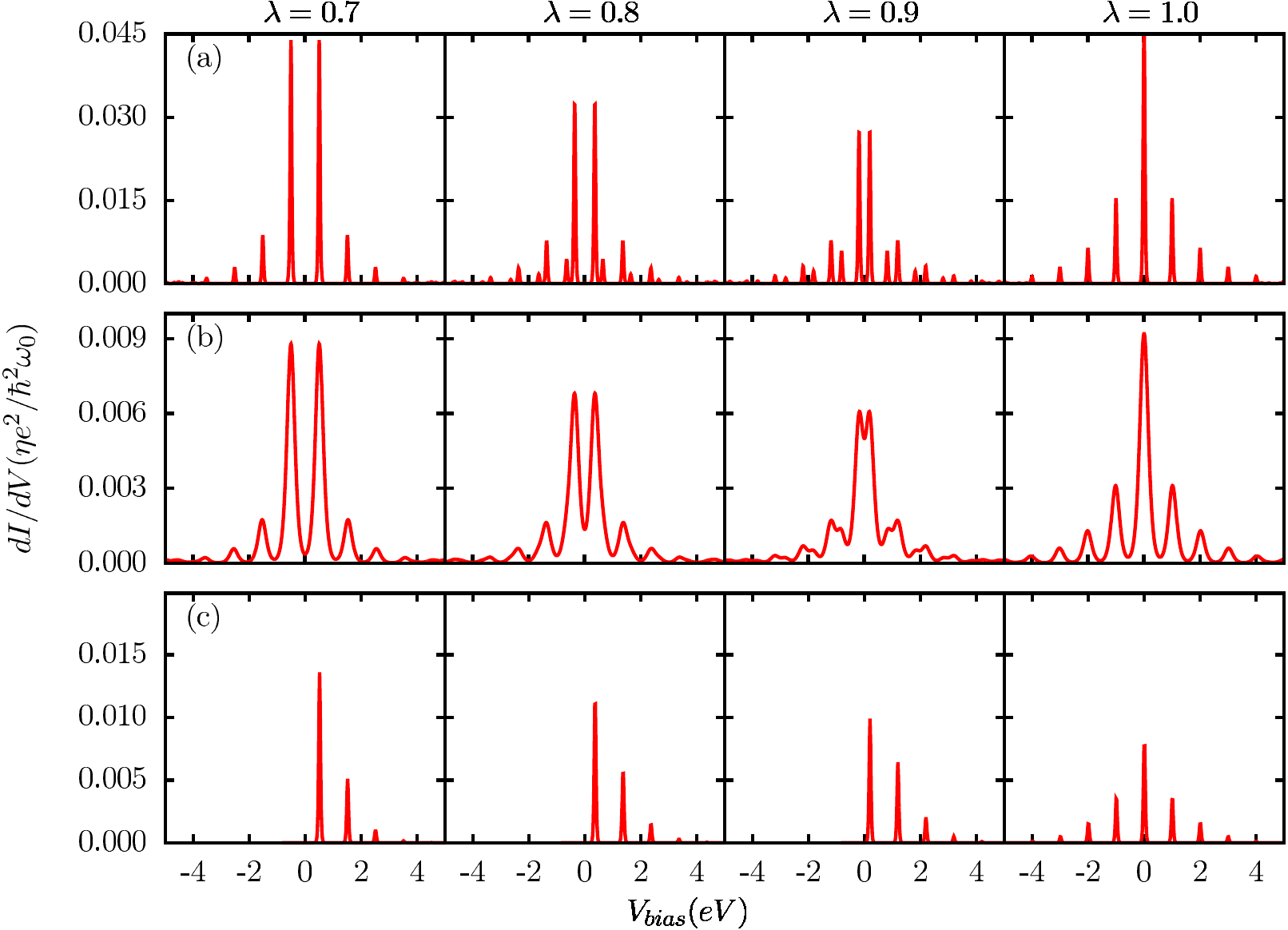}
\caption{\label{fig:sidebands}The phonon sidebands with different EPI strength
with (a) low temperature ($T=0.02\hbar\omega_0/k_B$) and symmetrically biased
voltage ($V_{L/R}=\pm V_{bias}$), (b) high temperature
($T=0.1\hbar\omega_0/k_B$) and symmetrically biased voltage, (c) low
temperature and asymmetric biased voltage ($V_L=V_{bias}$ and $V_R=0$). Other
parameters include $\varepsilon_0=1.0\hbar\omega_0$,
$\varepsilon_D=10\hbar\omega_0$ for all plots.}
\end{figure}

Besides the peak distances and peak width discussed above, other aspects
characterizing  the sidebands include the weights of the zero-phonon band and
the number of peaks. The investigation of these properties mainly focuses on the
effects of EPI strength, Fermi energy of the molecule, chemical potential and
temperature of the leads. In general, the higher order peaks will be suppressed
by the Frank-Condon factor \cite{Entin-wohlman2009,Hartle2011b, Monreal2010}. In
the framework of NEGF, Chen $et$ $al$. found that the weight from zero-phonon band
will decrease monotonically with increase of EPI strength and temperature
while the weights of higher order sidebands will increase and then decrease
\cite{Chen2005}. The chemical potentials of the leads will influence the
presence of the sidebands at both sides of the zero phonon peaks [Fig.~\ref{fig:sidebands}, panel (a) and (b)]. If one keeps
the chemical potential of one lead fixed and increases the chemical potential
of the other lead, the phonon sidebands will appear only at one side of the 0-th
order peak [Fig~\ref{fig:sidebands}, panel (c)]. However, if one fixes the Fermi-level of the molecule and changes the chemical potentials of both leads, phonon sidebands will appear at both sides\cite{Galperin2006, Chen2005}. We note that, in this section,
Fermi-level of the molecule is defined as $(\mu_L+\mu_R)/2$. 

In the framework of QME formalism described earlier, which is
exact in the weak system-lead coupling limit, we find phonon sidebands for the
single electronic level interacting with a single phonon mode. In this case the
zero-phonon peak will occur at the renormalized resonant level due to polaron
shift ($\varepsilon_0-\lambda^2/\hbar\omega_0$) and the sidebands will appear at
every distance of $\hbar\omega_0$. The peaks will appear at each side of
zero-phonon peak under symmetric change of the lead chemical potentials while
only appear at one side if we fix chemical potential of one lead. The EPI strength will not only shift the peaks, but also modulate the weights of the peaks. We find
that upon increasing either EPI strength or temperature, the weight of the
zero-phonon peak will decrease, which is consistent with the previous
work\cite{Chen2005}. However, interesting phenomena happen when the
renormalized level of the quantum dot gets close to the Fermi-level of the dot, 
[$\varepsilon_0-\lambda^2/\omega_0\approx (\mu_L+\mu_R)/2$], i.e., additional peaks
appear at each side. Those peaks have distance $\hbar\omega_0$ with the
zero-order peak at the opposite side. However, the two major peaks really merge
together, and those additional peaks disappear again. In the case of asymmetric change
of lead chemical potential (right most panel of Fig.~\ref{fig:sidebands} (c)),
we found peaks appear at both sides when the zero-phonon peaks merge together.

Another interesting perspective of the $I-V$ characteristics is the phenomenon
of the negative differential resistance (NDR), where the current decreases with
the increase of voltage bias. The NEGF formalism predicts that NDR is
impossible in ballistic electronic transport, but it will emerge in the
presence of EPI. NDR has been both theoretically
investigated\cite{Zazunov2006,Hartle2011b,Egger2008} and experimentally
measured\cite{LeRoy2004}. An important reason of NDR is due to the
redistribution of the molecular states. As discussed in the previous section,
the zero-phonon band carries major portion of electronic current. The
probability for the molecule to be in that state will be related to the
chemical potential of the leads. If one increases the bias via increasing the
chemical potential of one lead, one actually lifts the Fermi-level of the
molecule as well. If one brings the Fermi-level of molecule far away from the
eigenenergy of zero phonon state, the probability of the molecule to be in that
state will decrease and thus the current will decrease. Based on this analysis
one can draw several immediate conclusions: 1. If one increases the bias
simultaneously for both leads in pace and keeps the Fermi-level of the
molecule fixed, there will be no NDR. 2. If one treats the phonon fixed in the
equilibrium distribution, NDR will not appear\cite{Hartle2011b}. 3. If
the chemical-potential-varying lead couples stronger to the molecule than
the chemical-potential-fixed lead, the redistribution will be more sensitive,
thus the NDR will be enhanced.

\begin{figure}
\includegraphics[width=0.9\linewidth]{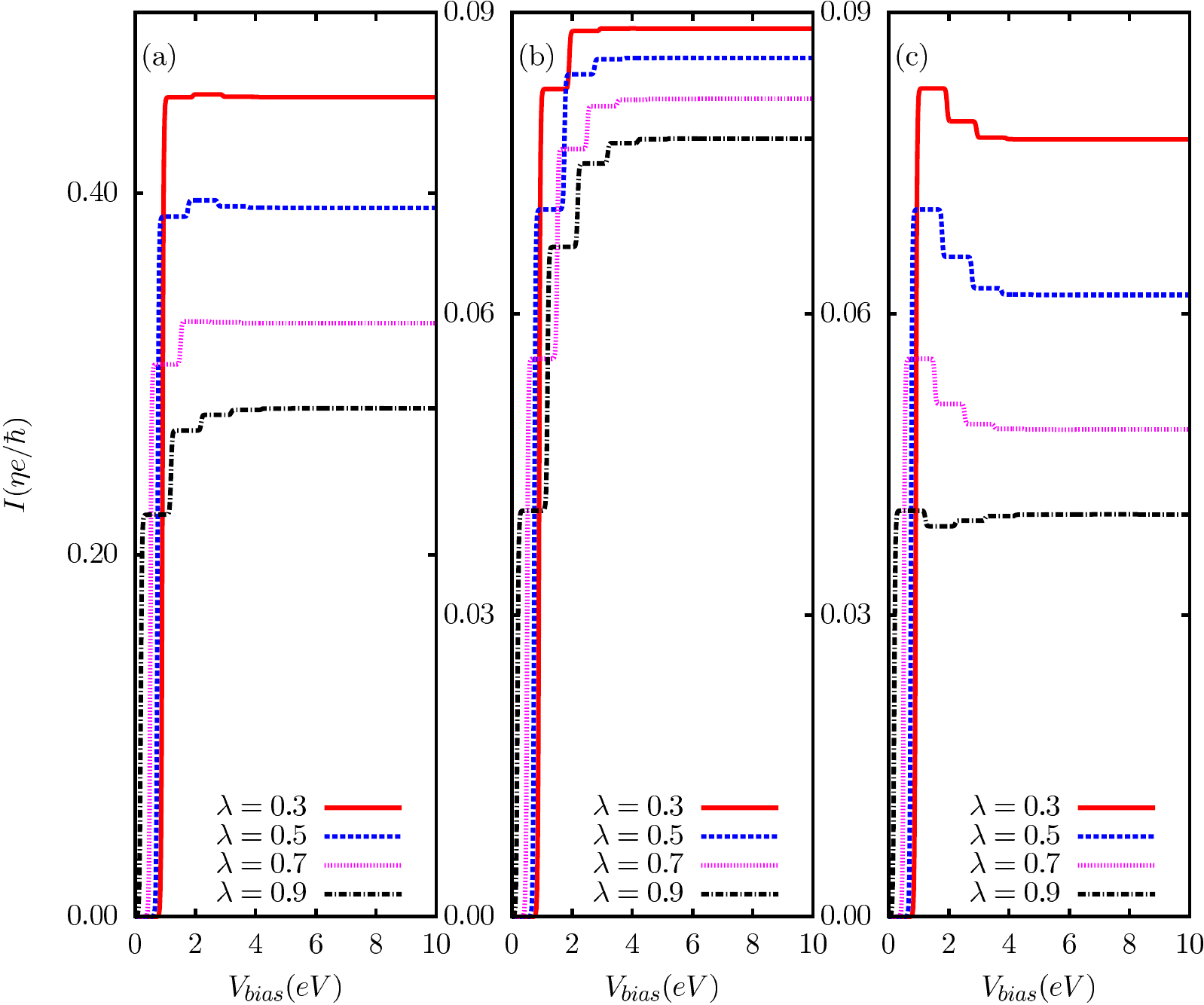}
\caption{\label{fig:NDR}Prediction of NDR using QME formalism with the coupling strength (a) $\eta_L=0.1\eta_R$, (b) $\eta_L=\eta_R$ and (c) $\eta_R=0.1\eta_L$. For all plots $V_R$ is fixed at 0 and $V_L=V_{bias}$. The temperature is fixed at $T=0.02\hbar\omega_0/k_B$ for all leads. Other parameters are the same as Fig.~\ref{fig:sidebands} }
\end{figure}

Figure~\ref{fig:NDR} shows the NDR predicted in the QME formalism. We find that
the NDR appears in the symmetric system-lead coupling. Moreover, NDR will be
enhanced if the chemical-potential-varying lead couples stronger to the
molecule than the other lead, but it will disappear in the other way around. We
also find that NDR is most pronounced in the moderate EPI regime, while
it is less significant in both weak and strong EPI regimes\cite{Hartle2011b}.

\subsection{Vibrational effect on thermoelectricity}
\label{subsec:qmethermoelectric}
In this part, we study the effects of EPI on the thermoelectric properties of
the nano-conductor.  We first look at the thermoelectric current, which is
the electronic current induced by a temperature difference between the leads.
Thermoelectric current exhibits quite different features comparing with voltage-bias current. It will increase monotonically and smoothly with the
increasing of the temperature bias. Therefore, there will be no phonon
sidebands. This is expected, because under temperature bias, the tunneling
channel of the electron is always restricted to the molecular state that is
close to the chemical potential of the leads. The increasing of temperature
will only excite more conducting electrons, but not be able to extend extra
tunneling channels. Therefore, there will be no sudden change of thermoelectric
current. Due to the same
reason, there will be no NDR effect as the increasing of the temperature bias
will always make more electrons to be involved in tunneling. The restriction on
tunneling channels also makes the thermoelectric current much smaller than the
voltage-bias current.

\begin{figure}
\includegraphics[width=0.9\linewidth]{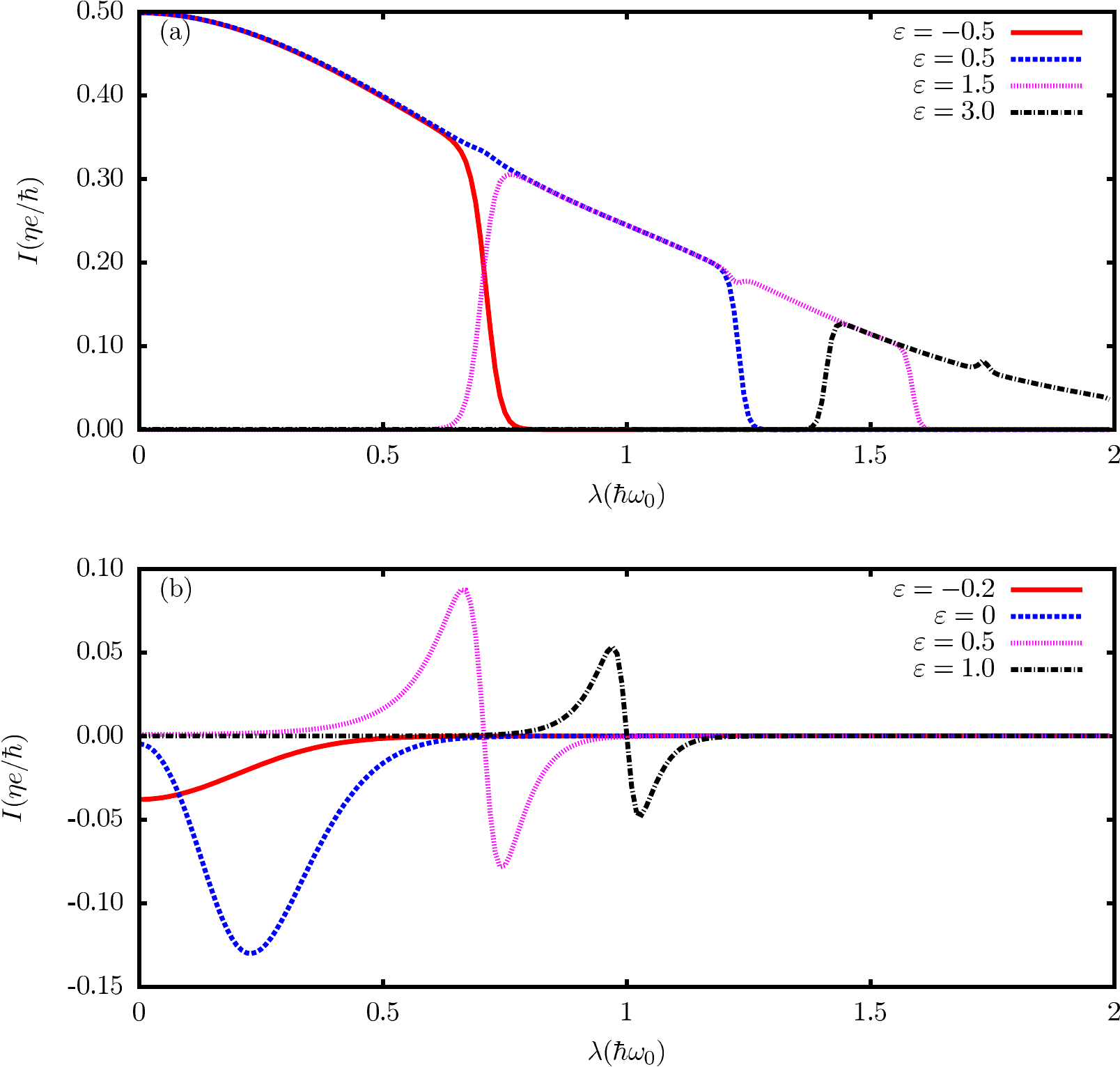}
\caption{\label{fig:current_lambda}The dependence of voltage-bias current (a) and thermoelectric current (b) on the EPI strength under different $\varepsilon_0$. For panel (a), $\mu_L=\hbar\omega_0$, $\mu_R=-\hbar\omega_0$ and $T^L=T^R=0.02\hbar\omega_0$. For panel (b), $T^L=0.08\hbar\omega_0$, $T^R=0.02\hbar\omega_0$ and $\mu_L=\mu_R=0$.}
\end{figure}
The dependence of electronic conductance on the chemical potentials of the leads
has been discussed in the Subsec.~\ref{subsec:loe}. The sign of the Seebeck
coefficient indicates that the currents will change direction with different
chemical potential. Another interesting perspective is to find the dependence
of currents on EPI strength. Figure \ref{fig:current_lambda} shows the plot of
voltage-bias current [panel (a)] and thermoelectric current [panel (b)] with
the EPI strength $\lambda$. For voltage-bias current, we find that the
current decays with EPI strength in general. More precisely, the maximum
current one can achieve via adjusting the $\varepsilon_0$ is decreasing
monotonically with EPI strength. This is due to the Frank-Condon blockage of
the current\cite{}. However, for each $\varepsilon_0$ we can find the enhancement of
current due to EPI, and that is mainly because of the polaron shift which can
bring the electron resonance level into the conduction band from outside. For
the thermoelectric current, the profile is quite different. We see that the
current can change sign with the increase of EPI strength when $\varepsilon_0$ is
higher than the Fermi level, which indicates that the EPI can switch the charge
carriers of the quantum dot between electrons and holes\cite{Zhou2015}. For
each $\varepsilon_0$ there exist two optimized values of EPI strength such that
the thermoelectric current maximizes. This optimized $\lambda$ shifts left
with decrease of $\varepsilon_0$ until disappear one by one at the $\lambda=0$
end.

\begin{figure}
\includegraphics[width=1.0\linewidth]{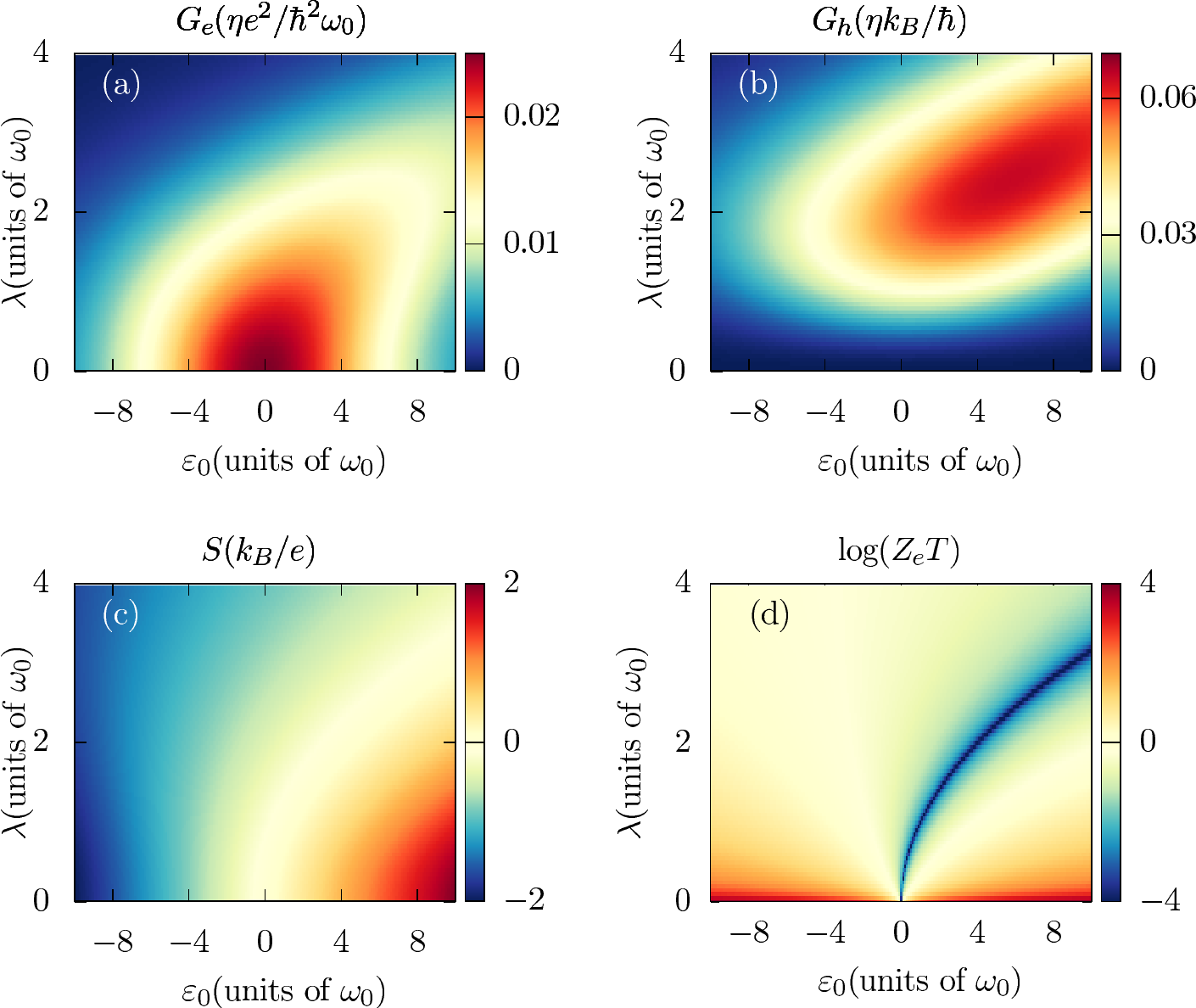}
\caption{\label{fig:thermoelectricity}The contour plot of electronic
conductance $G_{\rm e}$, thermal conductance $\kappaa$, Seebeck coefficient
$S$ and figure of merit $\ZT$ with respect to $\varepsilon_0$ and $\lambda$.
The parameters are:
$\mu = 0$, $T=5 \hbar\omega_0/k_B$. For this plot,  the phonon mode is coupled to its environments at temperature $T$ with coupling strength $\eta_E = 0.1 \eta$. Figure adapted with permission from Phys. Rev. B, {\bf 91}, 045410 (2015). Copyrighted by the American Physical Society.}
\end{figure}

One important quantity to describe the efficiency of the thermoelectric
material is the figure of merits $Z_{\rm e}T$. It is related to the
electronic conductance $G_{\rm e}$, Seebeck coefficient $S$, thermal
conductance $\kappaa$ via the formula $\ZT=G_{\rm e}S^2T/\kappaa$. Here we use
the notation $Z_{\rm e}T$ to denote the figure of merits of the system by
ignoring the thermal conductance due to phonons. So $\kappaa$ here only takes
account the thermal conductance due to electrons. The effect of phonons on the
figure of merit $\ZT$ is rather complicated, which is closely related to the
electron Fermi energy\cite{Koch2014,Perroni2014,Leijnse2010,Ren2012}, the
phonon energy \cite{Zianni2010}, the temperature and chemical potentials of the
leads\cite{Koch2014, Perroni2014,Ren2012,Hsu2012}.
Figure~\ref{fig:thermoelectricity} shows the dependence of the electronic
conductance, thermal conductance, Seebeck coefficient and figure of merit on
the electron energy $\varepsilon_0$ and EPI strength $\lambda$. The major
effect of EPI is on the thermal conductance of the molecular
junction\cite{Zianni2010}. The EPI can open extra channels, from which high
energy electrons can tunnel from hotter lead to colder lead, while low energy
electrons tunnel in the opposite direction. Therefore, the thermal conductance is
enhanced while the electronic conductance is not affected too much. Due to the
increase of the thermal conductance, $\ZT$ will be suppressed
drastically\cite{Zianni2010,Perroni2014}. Though the figure of merits $\ZT$
will be reduced quickly under the influence of phonon scattering in weak EPI
regime, it will gradually saturate at strong EPI regime. We
also would like to point out that when the electron energy is close to the
Fermi energy of the quantum dot, the Seebeck coefficient and hence the $\ZT$
will become very small. However, the phonon scattering can renormalized the
electron energy via polaron shift and thus the Seebeck coefficient can be
enhanced. As a result, EPI can enhance $\ZT$ in this particular parameter
regime.

\subsection{Local heating}
Local heating is an important phenomenon in molecular junction, not only due to
its own importance affecting the stability of the system, but also due to its close relation to the phonon sidebands\cite{MiAlMi04}, NDR and thermoelectric
effect\cite{Hsu2012,Hsu2010}. The distribution of the phonon states in
current-carrying system can be far away from equilibrium\cite{MiAlMi04}, or in some 
cases, may even lead to phonon instability\cite{JMP.2010,LuBrHe.2011,HaTh11,Simine13PCCP}. Phonons can be excited significantly by the voltage bias\cite{Hartle2011b}, and in turn affect the $I-V$ characteristics.
At each peak of the phonon sidebands, one can actually find a vibrational
excitation event\cite{Galperin2007, Hartle2011b}. Previous study also found
that the local heating can enhance the thermoelectric
efficiency\cite{Hsu2010,Hsu2012}. However, in most cases local heating is not
preferred because it will affect the stability of the system\cite{JiangJW2011joule}
and introduce noise to the measurements\cite{Blanter2004,Clerk05,Mozyrsky04,Zippilli10}. Therefore, lot of effort has been put into
cooling the system using electronic current, such as using superconducting single
electron transistor\cite{Naik06}, or double quantum dots\cite{Zippilli10}. 

\begin{figure}
\includegraphics[width=0.9\linewidth]{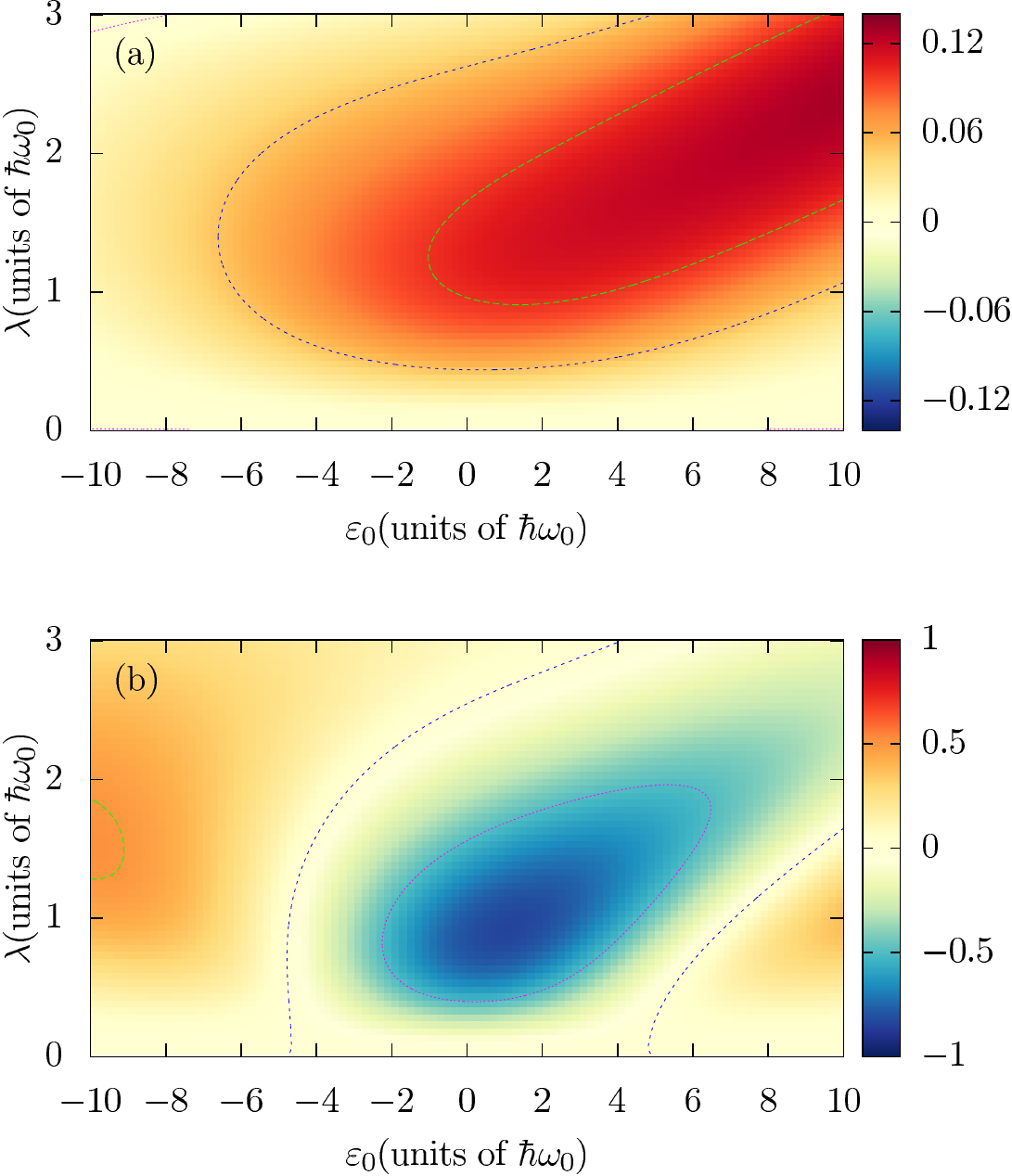}
\caption{\label{fig:localheating} The correction of phonon number due to
electronic current under voltage bias (a) and temperature bias (b). The
parameters are: $\mu_L = -\mu_R = 2\hbar\omega_0$, $T^L=T+\Delta T$,
$T^R=T-\Delta T$, $\Delta T=3\hbar\omega_0/k_B$ and $T = 5\hbar\omega_0/k_B$.
The phonon is weakly coupled to its own environment at temperature $T$.  Figure
adapted with permission from Phys. Rev. B, {\bf 91}, 045410 (2015). Copyrighted
by the American Physical Society.}
\end{figure}

Here we study the effects of the electronic current on the phonon mode of the nano-conductor. To investigate the local heating, effective temperatures are usually defined in
various ways\cite{DuDi11,GaRaNi.2007,Lu07,Dubi09E,Jacquet12,Bergfield13}. In this part, since we only have one single phonon mode, instead of
defining the effective temperature, we use average phonon number to characterize
the local heating effect. The way to specify the electronic current induced
heating is to compare the nonequilibrium phonon number $n_{\rm neq}$ with
equilibrium phonon number $n_{\rm eq}$. When the molecule is weakly coupled to the
leads, the molecular states statistics will follow canonical distribution
$\rrho=e^{-\beta H_S}/\mathrm{Tr}(e^{-\beta H_S})$ and thus the equilibrium
phonon number can be calculated exactly as \cite{Zhou2015}
\begin{equation}
\label{eq:polaron}
n_{\rm eq}=\frac{1}{e^{\beta_{}\hbar\omega_0}-1}+\frac{\lambda^2/(\hbar\omega_0)^2}{e^{\beta_{}(\varepsilon_0-\lambda^2/(\hbar\omega_0))}+1}.
\end{equation}
The first term is the Bose-Einstein distribution function, the second term
is a correction due to the polaron energy shift.
The nonequilibrium phonon number can be calculated from the nonequilibrium
reduced density matrix obtained from the QME.
Figure~\ref{fig:localheating} shows the difference of phonon numbers under
voltage bias (top) and temperature bias (down). For voltage bias,
$\Delta n$ is always positive which indicates that the system is always heated
up.  The yellow regime where  $\Delta n\approx 0$  is the regime where the
electronic current vanishes. When there is electronic current passing
through, the heating effect is generally more pronounced in stronger EPI 
regime. However, for the temperature bias, we find both heating
($\Delta n>0$) and cooling regimes ($\Delta n<0$). Therefore, the local
heating effect is not only related to the magnitude of electronic current, but
also related to the energy each electron carries when it tunnels into
the molecule. For the thermoelectric current, low energy electron can tunnel
from the cooler lead to the molecule, absorb a phonon and tunnel to
the hotter lead, resulting in cooling of the molecule. Such process is
impossible in the voltage-bias case in the present setup as the electron is
always flowing from the higher chemical potential side to the lower chemical
potential side.
\section{Current-induced semi-classical Langevin dynamics}
\label{sec:gle}
In the previous two sections, we mainly look at the effect of phonons on the
electric and thermoelectric transport properties of electrons, which is also
the focus of most published work.  But, to study the current-induced forces,
and their effect on atomic dynamics, we need to turn around. In this section,
starting from the total Hamiltonian $H_{\rm tot}$, we derive a semi-classical
Langevin equation to describe the atomic dynamics of the system, coupled to
both phonon and (nonequilibrium) electron leads. The Langevin equation applies
to the weak EPI regime, since similar to Sec.~\ref{sec:negf} our expansion
parameter is the interaction matrix $M$. However, the derivation is not limited
to the form of $H_{\rm tot}$. Following the same procedure we discuss below, we
can also do an adiabatic expansion of the electron influence functional over
the velocity of the ions. In that case, the Langevin equation applies to slow
ions, but the EPI could be of arbitrary
magnitude\cite{JMP.2010,luprb12,BoKuEgVo.2011,BoKuEg.2012}.

The advantage of the Langevin equation approach is that, we can easily include
the anharmonic phonon interaction, as in other molecular dynamics method. The
anharmonic interaction is crucial in dealing with current-induced dynamics.
This is because the phonon modes that interact with electrons are normally
high-frequency ones, while the low frequency modes conduct heat from the system
to surrounding electrodes more effectively. The energy transfer from high to
low frequency modes is possible only when anharmonic interaction is included.
Although possible, it is not a trivial task to incorporate anharmonic
interactions in the NEGF or QME approach.

\subsection{Initial states and reduced density matrices}

We assume at a remote past $t_0$ and earlier time, the central system is
decoupled with the leads and there is also no EPI, so that the electrons and phonons
are also decoupled. The density matrix is assumed to be product of known
equilibrium states.  For example, the left lead is specified by
\begin{eqnarray}
\rho^L_{\rm e} &\propto& e^{-\beta_L(H^L_{\rm e} - \mu_L N^L)}, \\
\rho^L_{\rm p} &\propto& e^{-\beta_L H_{\rm p}^L}.
\end{eqnarray}
Exactly what to take for the center is not important as for steady
state with $t_0 \to -\infty$, the results do not depend on it (except
maybe in very subtle cases).

The density matrix (of the whole system) is governed by the von
Neumann equation, and formally we can write
\begin{equation}
\rho(t) = U(t,t_0) \rho(t_0) U(t_0, t),
\end{equation}
where 
\begin{equation}
U(t,t_0) = T e^{-(i/\hbar) \int_{t_0}^t H_{\rm tot}(t') dt'}
\end{equation}
assuming $t> t_0$.  For the other case of $t<t_0$, the time-order
operator $T$ should be replaced by the anti-time order operator.  We are
interested only in the center, so the leads degrees of freedom will be
traced out.  For notational simplicity, we assume only one left lead.
The result for two or more leads is trivially generalized.  We define
\begin{eqnarray}
\tilde \rho(t) &=&  {\rm Tr}_L \rho(t), \\
\tilde \rho_{{\rm p}} (t) &=&  {\rm Tr}_{{\rm e}} \tilde \rho(t), 
\end{eqnarray}
where the first reduced density matrix only eliminates the lead, while
the last one eliminates the electrons as well,
leaving only an effective density matrix for the phonons.  The procedure
to eliminate the lead for both the electrons and phonons
follows the standard method of Feynman and Vernon
\cite{FEYNMAN1963}, except that we need to be careful for the
electrons which are fermions.  Since there is no coupling between
electrons and phonons in the leads, the phonon and
electron degrees of freedom can be done separately (the initial
density matrix is a product of the two).

\begin{figure}
\includegraphics[width=\columnwidth]{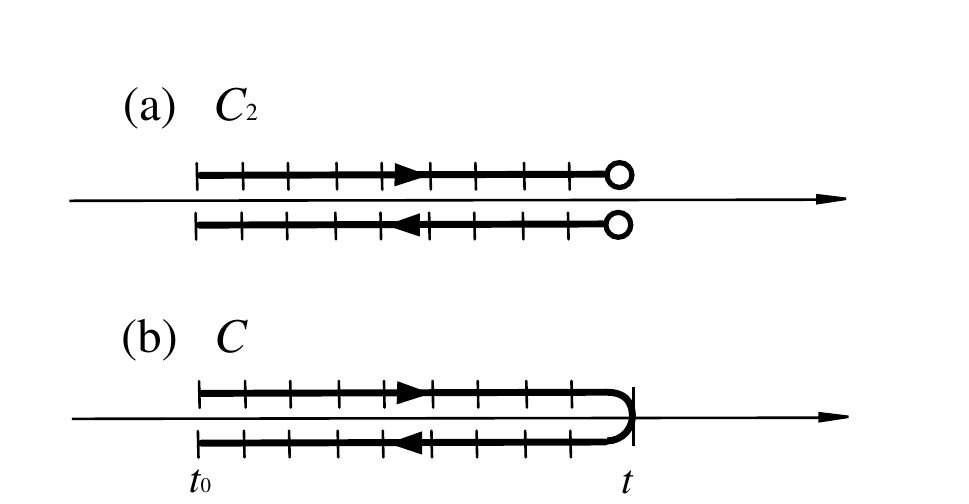}%
\caption{\label{fig-Cpath}Two types of paths in the Feynman-path
integrals.  (a) two-segment path $C_2$. (b) Keldysh type closed
contour $C$.  The ticks represent the discretized integration
variables; the open circles are not integrated.}
\end{figure}

\subsection{Influence functional for phonons}
The matrix elements of the density operator $\rho(t)$ is taken in the
basis of the coordinates $u$.  Following the standard treatment
\cite{kleinert,weiss}, the density matrix is then given by
\begin{equation}
\label{eq-urho}
\langle u' |\rho(t) |u \rangle \propto 
\int \mathcal{D}[u] e^{(i/\hbar) \int_{C_2} \mathcal{L}(u, \dot{u}) d\tau}
\rho_0(u'_0, u_0).
\end{equation}
The path $C_2$ consists of two segments (see Fig.~\ref{fig-Cpath}(a))
running from time $t$ with coordinate variable $u$ back to $t_0$ at
variable $u_0$, and then second segment running from $t_0$ with
variable $u'_0$ forward to time $t$ with variable $u'$.  Since the
arbitrary constant in the proportionality can be fixed by
normalization (${\rm Tr} \rho = 1$), we need not specify precisely the
measure associated with $\mathcal{D}[u]$.  The path integral integrates
all the intermediate variables except the two open ends at time $t$.

To eliminate the lead, the phonon Lagrangian is split into $\mathcal{L} =
\mathcal{L}^C + \mathcal{L}^L - V$ where $V$ is the coupling potential
energy term between the lead and center, given by $V = (u^C)^T V^{CL}
u^L$.  The integration volume elements are also split $\mathcal{D}[u]
= \mathcal{D}[u^C] \mathcal{D}[u^L]$.  The initial distribution is
assumed product type $\rho_0^L \otimes \rho_0^C$.  Taking the trace of
Eq.~(\ref{eq-urho}) means that we identify $u^L$ and $(u')^L$ as the
same variable $u^L$ and integrate it out.  As far as variable $u^L$ is
concerned, the path is of Keldysh type, i.e., running from $t_0$ to
$t$ from above and then back from $t$ to $t_0$, as shown in
Fig.~\ref{fig-Cpath}(b).  The reduced density matrix is then
\begin{equation}
\tilde \rho(t) = \int \mathcal{D}[u^C] e^{(i/\hbar) \int_{C_2} 
\mathcal{L}^C d\tau} I[u^C(\tau)] \rho_0^C((u')_0^C, u_0^C),
\end{equation}
with the influence functional
\begin{equation}
I[u^C(\tau)] = \int \mathcal{D}[u^L] \rho_0^L((u')_0^L, u_0^L) 
e^{(i/\hbar) \int_C(\mathcal{L}^L - V) d\tau}.
\end{equation}
The above expression can be further simplified.  Firstly, the initial
distribution of the lead is in thermal equilibrium, $\rho_0^L \propto
e^{-\beta_L H_L}$. Secondly, we can rewrite the path integral formula
back into the operator form; thus, we obtain
\begin{eqnarray}
I[u^C(\tau)] &=& {\rm Tr} \left[ \frac{e^{-\beta_L H^L}}{Z_L} T_C 
e^{-(i/\hbar) \int_C(H^L + V) d\tau} \right]  \nonumber \\
          &=& \langle T_C 
e^{-(i/\hbar) \int_C(H^L + V) d\tau} \rangle_{{\rm eq.L}},
\end{eqnarray}
where $Z_L$ is the canonical partition function, $T_C$ is contour order
operator, the subscript eq.L stands for equilibrium average with
respect to the left lead.  One more transformation can be made to
simplify it further.  The time-dependence (or rather, contour time $\tau$
dependence) is understood to be in the Heisenberg picture governed by
the Hamiltonian $H^L + V(\tau)$ which is different in the forward and
backward direction.  We can work in the interaction picture, thus
eliminate the explicit $H^L$ from the formula.  The resulting equation is
\begin{equation}
I[u^C(t)] = 
\langle T_C e^{-(i/\hbar) \int_C V_I(\tau) d\tau} \rangle_{{\rm eq.L}}.
\end{equation}
This is the same equation [Eq.~(5)] given in Ref.~\onlinecite{stockbuger02}.

The influence functional can be calculated explicitly since the
interaction $V$ is a quadratic form, and the contour operator
naturally leads to contour ordered Green's functions and Wick's theorem
is valid.  We define the contour ordered phonon Green's function of
the lead as
\begin{equation}
d_L(\tau, \tau') \equiv - \frac{i}{\hbar} 
\langle T_C u^L(\tau) u^L(\tau')^T \rangle_{{\rm eq.L}}.
\end{equation}
Expanding the exponential (or a cumulant expansion, expanding and taking logarithm), we get
\begin{widetext}
\begin{eqnarray}
&&I[u^C(t)] = \bigl\langle T_C \bigl( 1 -  \frac{i}{\hbar} \int_C V_I(\tau) d\tau 
             - \frac{1}{2\hbar^2} \int_C d\tau \int_C d\tau' V_I(\tau) V_I(\tau') + \cdots \bigr)\bigr\rangle\nonumber \\
&=& T_C \bigl(1 - \frac{1}{2\hbar^2}\int_C\int_C u^C(\tau)^T V^{CL} \langle 
T_C u^L(\tau) u^L(\tau')^T \rangle V^{LC} u^C(\tau') d\tau d\tau' + \cdots \bigr) \nonumber \\
&=& T_C \bigl(1 - \frac{i}{\hbar} \frac{1}{2} \int_C d\tau\int_C d\tau' 
u^C(\tau)^T \Pi_L(\tau, \tau') u^C(\tau') + \cdots \bigr) \qquad\qquad\qquad\qquad \nonumber \\
&=&e^{-\frac{i}{\hbar} \frac{1}{2} \int_C d\tau\int_C d\tau' 
u^C(\tau)^T \Pi_L(\tau, \tau') u^C(\tau')}.\qquad\qquad\qquad\qquad
\label{eq-pi-influ}
\end{eqnarray}
\end{widetext}
The first order term (and all odd order in $V$ terms) vanishes, because $\langle u^L \rangle = 0$.  We
have defined the ubiquitous contour ordered self-energy due to the
lead as
\begin{equation}
\Pi_L(\tau, \tau') = V^{CL} d_L(\tau, \tau') V^{LC}.
\end{equation}
It can be shown that the last line in the above derivation is an exact
result.

It is instructive to clarify and compare with other notations used in
the literature. A commonly used notation is (e.g.,
Ref.~\onlinecite{weiss})
\begin{eqnarray}
\ln I &=& -\frac{1}{\hbar} \int_{t_0}^t dt_1 \int_{t_0}^{t_1} dt_2 
 \left( u^{+}(t_1) - u^{-}(t_1) \right)^{T} \nonumber \\ 
&&\times \left[ L(t_1, t_2) u^{+}(t_2) - L^{*}(t_1, t_2) u^{-}(t_2) \right].
\label{eq-weiss}
\end{eqnarray}
Using the rules $\int_{C} d\tau \to \sum_{\sigma}\int_{t_0}^t \sigma
dt$, and $\Pi(\tau, \tau') \to \Pi^{\sigma, \sigma'}(t,t')$,
$u^C(\tau) \to u^{\sigma}(t)$ where $\sigma = +$ or $-$ for the upper
and lower branch, the relations among the various self-energies,
and symmetry relation $\Pi_{ij}^{>}(t,t') = \Pi_{ji}^{<}(t',t)$, we
can rewrite Eq.~(\ref{eq-pi-influ}) in the form of
Eq.~(\ref{eq-weiss}).  By comparison, we find
\begin{equation}
L(t,t') = i \Pi^{>}_L(t,t'),\quad
L^{*}(t,t') = i \Pi^{<}_L(t,t').\quad
\end{equation}
$\alpha(t,t') \equiv L(t,t')$ is the notation used by Schmid
\cite{SC.1982}.

\subsection{Influence functional for electrons}
The derivation of the influence functional for the electrons is
similar except that we have to deal with grassmann integrals\cite{PerNotes,Chang85,HedegardCaldeira,Chen87}. We
follow the approach of Weinberg \cite{weinberg-book}.  To do the trace
over the leads we need a specific representations for the operator
$\rho$. For the electrons, we use the coherent state characterized by
grassmann numbers such that
\begin{equation}
\hat{c}_i |c \rangle = c_i |c \rangle.
\end{equation}
The hat denotes operator, without hat, it is a grassmann number.  The
state $|c \rangle$ has explicit form given as
\begin{eqnarray}
| c\rangle &\equiv&  e^{ \sum_{i} \hat{c}_i^\dagger c_i } | 0 \rangle,\\
\langle c | &\equiv&  \langle 0 | \prod_i \hat{c}_i e^{ \sum_{i} c_i \hat{c}_i^\dagger}.
\end{eqnarray}
The orthogonality is in the form
\begin{equation}
\label{eq-prod-i}
\langle c ' | c \rangle = \prod_i (c_i -c'_i).
\end{equation}
Similar states for the creation operator $\hat{c}_j^\dagger$ can be
constructed with eigenvalue (a grassmann number) $\tilde{c}$, given
the following results (similar to inner product of eigen states of $u$
and its conjugate momentum $p$ and completeness of the eigenstates).
\begin{eqnarray}
\langle c | \tilde c \rangle &=& \cdots e^{-\sum_{j} \tilde c_j c_j}, \\
\langle \tilde c | c \rangle &=& e^{+\sum_{j} \tilde c_j c_j},\\
1 &=& \int | c \rangle \tilde{\prod}_j (-dc_j) \langle c |, \\
1 &=& \int | \tilde c \rangle \tilde{\prod}_j (dc_j) \langle \tilde c |.
\end{eqnarray}
The $\cdots$ is an extra $+$ or $-$ sign factor which we'll not keep
track.  The tilde over the product sign means that order is in exactly
the opposite canonical order [e.g., that of Eq.~(\ref{eq-prod-i})].
With the above very sketching outline, the fermion evolution operator
\begin{equation}
	U(t,t_0) = T e^{- (i/\hbar) \int_{t_0}^t H_{}(\tau) d\tau}
\end{equation}
can be represented as a path integral of the form
\begin{equation}
\int D[\tilde c,c] e^{iS_{\rm e}/\hbar}, 
\end{equation}
with the action
\begin{equation}
S_{\rm e} = - \int d\tau \left( {\tilde c}^T H c - i\hbar {\tilde c}^T \frac{\partial c}{\partial \tau} \right).
\end{equation}
The lead influence functional can be then obtained with the same
procedure as that for the phonons.  The result involves an integral
kernel which is exactly the contour ordered self-energy of the lead
$\Sigma_L(\tau, \tau')$.

\subsection{Reduced density matrix for phonon in center}
Putting things together, the reduced density matrix, when the leads
are eliminated, has the form
\begin{eqnarray}
\langle (u')^C, (\tilde c')^C, (c')^C|\tilde \rho(t) |
u^C, \tilde{c}^C, c^C \rangle \propto \qquad\qquad\qquad\qquad \nonumber \\
\int \mathcal{D}[u^C, \tilde{c}^C, c^C] e^{i\,S/\hbar} \rho_0(
(u'_0)^C, (\tilde c'_0)^C, (c'_0)^C;
u_0^C, \tilde{c_0}^C, c_0^C) \nonumber
\end{eqnarray}
where 
\begin{eqnarray}
S &=& S_{\rm p}^C + S_{\rm e}^C + S_{\rm ep}^C + S_{\rm p}^I + S_{\rm e}^I,\\
S^C_{\rm p} &=& \int_{{C_2}}d\tau \left(\frac{1}{2} \dot{u}^2 - \frac{1}{2}u^T K^Cu- V_n(u) \right),\\
S^C_{\rm e} &=& \int_{{C_2}}d\tau \left( i\hbar \tilde{c}^T 
\frac{\partial c}{\partial \tau} - \tilde{c}^T H^C c \right),\\
S_{\rm ep}^C &=& - \int_{C_2} d\tau \sum_{k} u_k \tilde{c}^T M^k c,\\
S_{\rm p}^{I} &=& - \frac{1}{2} \int_{C_2} d\tau \int_{C_2} d\tau'
u(\tau)^T \Pi_L(\tau, \tau') u(\tau'),\\
S_{\rm e}^{I} &=& - \int_{C_2} d\tau \int_{C_2} d\tau'
\tilde{c}(\tau)^T \Sigma_L(\tau, \tau') c(\tau').
\end{eqnarray}
The ordinary number (column vector) $u$ and grassmann number $\tilde
c$, $c$ involve only the degrees of freedom of the center.  For
notational simplicity, we have dropped the superscript $C$.  Note that
the electron terms do not have the characteristic factor $1/2$ as
$\tilde{c}$ and $c$ are independent variables.

The dependence on $\tilde{c}$ and $c$ is a bi-linear form, thus the
path integral over them can be done analytically.  This gives the
reduced density matrix of the phonon only, as
\begin{equation}
\label{eq-Sp}
\langle u' | \rho_{\rm p}(t) | u \rangle \propto
\int \mathcal{D}[u] e^{(i/\hbar)(S^C_{\rm p} + S_{\rm p}^I)} I_p \rho_{0,{\rm p}}.
\end{equation}
The influence functional to the phonons due to electrons is given by
\begin{equation}
I_p \propto {\rm det}\left( \delta(\tau, \tau') \left\{
I i \hbar \frac{\partial }{\partial \tau} - H^C - u_k(\tau) M^k \right\}
- \Sigma_L(\tau, \tau') \right).
\end{equation}
Interpreting the $\tau$ in the above as Keldysh variable defined on
$C$ has a problem.  As agreed, the contour is supposed to be on $C_2$
with the $t_0$ end connected with the initial distribution of the
electrons $\rho_0$ at the center.  However, if we assume that in the
limit $t_0 \to -\infty$ the results should not depend on the
distribution of the center, we can ignore this initial distribution
and it is completely fixed by the lead.  But we cannot give a
mathematically sound justification here.

Similar to that in Ref.~\onlinecite{weinberg-book} for the field
theory of quantum electrodynamics, we want to put the influence
functional in an exponential form.  This can be done using the
formula, ${\rm Det}(A) = e^{{\rm Tr}{\rm ln} A}$, and the expansion of the
function $\ln(1 + x) = x - x^2/2 + \cdots$ for small $x$, given
\begin{equation}
\label{eq-Ip}
I_p = {\rm det}(G_0^{-1} + y) = {\rm det}(G_0^{-1}) 
\exp\left( \sum_{n=1}^{\infty} \frac{(-1)^{n+1}}{n} {\rm Tr}[(G_0 y)^n] \right),
\end{equation}
where 
\begin{eqnarray}
G_0^{-1} &=& \delta(\tau, \tau') \left[
I i \hbar \frac{\partial }{\partial \tau} - H^C  \right]
- \Sigma_L(\tau, \tau'),\\
y &=& -\delta(\tau, \tau') \sum_{k} u_k(\tau) M^k.
\end{eqnarray}
$G_0^{-1}$ and $y$ are matrices indexed by lattice sites $j$ as well as
contour time $\tau$.  And, if the proper metric for a
discretization of the time is chosen so that ${\det}(G_0^{-1})$ can be
meaningful, we can identify $G_0$ as the electron contour
ordered Green's function when there is no EPI
defined in Sec.~\ref{sec:negf}.
Since ${\det}(G_0^{-1})$ is independent of $u$, the effective action for the
phonon is only determined by the exponential factor, which is a
polynomial (functional) in $u$.  With some caveat regarding the
initial distribution, Eqs.~(\ref{eq-Sp}) and (\ref{eq-Ip}) offer a
formally exact solution to the problem.

The first two terms, written out explicitly in terms of the contour
ordered Green's functions are
\begin{eqnarray}
{\rm Tr}\left[G_0 y\right] = - \sum_{k} {\rm Tr}\left[ G_0(\tau, \tau) M^k \right] u_k(\tau),\qquad\qquad\\
-\frac{1}{2} {\rm Tr}\left[ (G_0 y)^2 \right] = 
-\frac{1}{2} \sum_{j,l,m,n} \int d\tau d\tau' d\tau'' d\tau'''
{G_0}_{jl}(\tau, \tau') \nonumber \\
y_{lm}(\tau', \tau'') {G_0}_{mn}(\tau'', \tau''') y_{mj}(\tau''',\tau) \nonumber \\
= - \frac{i}{2\hbar}\int d\tau \int d\tau' u(\tau)^T \Pi_{\rm ep}(\tau, \tau') u(\tau'),\qquad\qquad
\end{eqnarray}
with
\begin{equation}
\Pi_{\rm ep}^{kk'}(\tau,\tau') =  -i\hbar {\rm Tr} \left[ G_0(\tau', \tau) M^k G_0(\tau, \tau') M^{k'} 
\right].
\end{equation}

\subsection{Semi-classical approximation}
If we ignore the linear term in $u$ which produces a constant force,
the effect of which is to shift the equilibrium positions, and also
neglect higher order contributions, we end up with a quadratic form
for the effective action
\begin{eqnarray}
S_{\rm eff} &=& \int_{C_2} d\tau \left( \frac{1}{2} \dot{u}^2 - 
\frac{1}{2} u^T K^C u - V_n(u) \right) \nonumber \\
&& - \frac{1}{2} \int_{C_2} d\tau \int_{C_2} d\tau'
u(\tau)^T \Pi_{\rm tot}(\tau, \tau') u(\tau').
\end{eqnarray} 
with $\Pi_{\rm tot} = \Pi_L + \Pi_R + \Pi_{\rm ep}$ as in Eq.~(\ref{eq:pitot}).
We have also included the right phonon lead.
A generalized Langevin equation can be derived from the above action\cite{SC.1982,luprb12}
\begin{equation}
\label{eq:gle}
\ddot{u} = - K^C u + F_n - \int^{t} \Pi_{\rm tot}^r(t-t') u(t') dt' + \xi,
\end{equation}
where $F_n = - \partial V_n/\partial u$, $\Pi_{\rm tot}^r$ is the retarded total self-energy, and the noises
satisfy
\begin{eqnarray}
\langle \xi(t) \rangle &=& 0,\\
\langle \xi(t) \xi^T(t') \rangle &=& i\hbar \frac{1}{2} \left( \Pi_{\rm tot}^{>}(t-t') 
+ \Pi_{\rm tot}^{<}(t-t') \right) \nonumber  \\
&=& i\hbar \bar{\Pi}_{\rm tot}(t-t').
\end{eqnarray}
We note that the effect of the electron leads to the phonons 
has exactly the same form as that of the phonon
leads. The self-energy consists of a sum of
contributions of the two sources.

\subsection{Applications}
Before discussing the applications, we note that similar generalized Langevin equation as
Eq.~(\ref{eq:gle}) can be derived by doing an adiabatic expansion over the
momenta of the ions to the 2nd order\cite{JMP.2010,HuMeZeBr.2010,luprb12,BoKuEgVo.2011}.
These equations have been used in different
perspective\cite{wang_quantum_2007,Lu08JPCM,WaNiJi.2009,Ceriotti09,Dammak09,Ceriotti12,Bedoya14,kantorovich_generalized_2008,kantorovich_generalized_2008-1,BrHe95,BrHe.1994a,HETU.1995,JMP.2010,HuMeZeBr.2010,lu2011,LuBrHe.2011,luprb12,Lu15,Dzhioev2011,BoKuEgVo.2011,BoKuEg.2012,Stella14,Ceriotti11}.
Its most important feature is the inclusion of the quantum nature of the electron
and phonon leads. For example, the zero point motions of atoms are correctly taken into account,
and proved critical in determining the thermal and structural properties of materials made from
light elements\cite{Ceriotti09,Dammak09,Ceriotti12,Bedoya14,kantorovich_generalized_2008,kantorovich_generalized_2008-1,Stella14,Ceriotti11}.
Including the correct Bose distribution of the phonons opens a way to study the quantum ballistic phonon transport by doing classical molecular dynamics. In Refs.~\onlinecite{wang_quantum_2007} and
\onlinecite{WaNiJi.2009} the transition from ballistic to diffusive phonon
thermal transport is studied using this approach. In
Refs.~\onlinecite{JMP.2010} and \onlinecite{Lu08JPCM}, including the
nonequilibrium electrons, current-induced dynamics have been studied.  Several
interesting effects have been predicted or confirmed, and their effects on the
stability of the system  are studied. For example, it has been shown that (1)
the current-induced forces are not conservative\cite{DuMcTo.2009}, (2) the
atoms feel an effective magnetic force, originating from the Berry phase of the
electrons\cite{JMP.2010}. Moreover, the power of the Langevin approach is
to be able to include the anharmonic phonon-phonon interactions classically,
and treat the EPI quantum-mechanically.  This enables one to study the energy
transport between different phonon modes, between electrons and phonons at the
same time. The exploration of its power and potential is still under way.

As an example, we consider the heat generation in a $4\times 2$ graphene armchair
ribbon, see Ref.~\onlinecite{WaNiJi.2009} for the definition of structure parameters. 
The electronic structure and EPI matrix are obtained from a combined
SIESTA\cite{Soler.02} + TranSIESTA\cite{BrMoOr.2002} + Inelastica\cite{FrPaBr.2007} calculation, while the Brenner potential is used
for the inter-atomic interaction. To reduce the simulation time, we have ignored
the energy-dependence of the electronic structure. As a result, the electronic friction becomes
time-local. The Langevin equation becomes (after an integration by part)
\begin{equation}
\ddot{u} = F^C - \int^t \Gamma(t-t') \dot{u}(t') dt' - 
\hbar \eta \dot{u} - eV \xi^{-} u  + \xi  + f,
\end{equation}
where $F^C$ is the force from the second-generation Brenner potential,
$d \Gamma(t) /dt = \Pi^r(t)$ is the phonon retarded self-energy due to
two leads, $V$ is applied bias voltage.  The $eV \xi^{-}u$ term gives
a nonconservative force as $\xi^{-}$ is antisymmetric.  $\xi = \xi_L +
\xi_R$ is the noise due to left and right phonon leads, while $f$ is
the noise due to electron bath. The expression for the electronic friction and noise correlation is the same as Eqs.~(56-63) in Ref.~\onlinecite{luprb12}.  Further
implementation details can be found in Ref.~\onlinecite{WaNiJi.2009}.
The phonon heat current is calculated using 
\begin{eqnarray}
I_{{\rm p}}^\alpha &=& \left\langle \dot{u}^T \left[ 
-\int^t \Gamma_\alpha(t-t') \dot{u}(t') dt' + \xi_\alpha \right]\right\rangle,
\end{eqnarray}
where $\alpha= L, R$.
In steady state, the energy flow balances, and the heat generation is calculated
according to $Q=-I_{\rm p}^L-I_{\rm p}^R$.

\begin{figure}[t]
\includegraphics[width=\columnwidth]{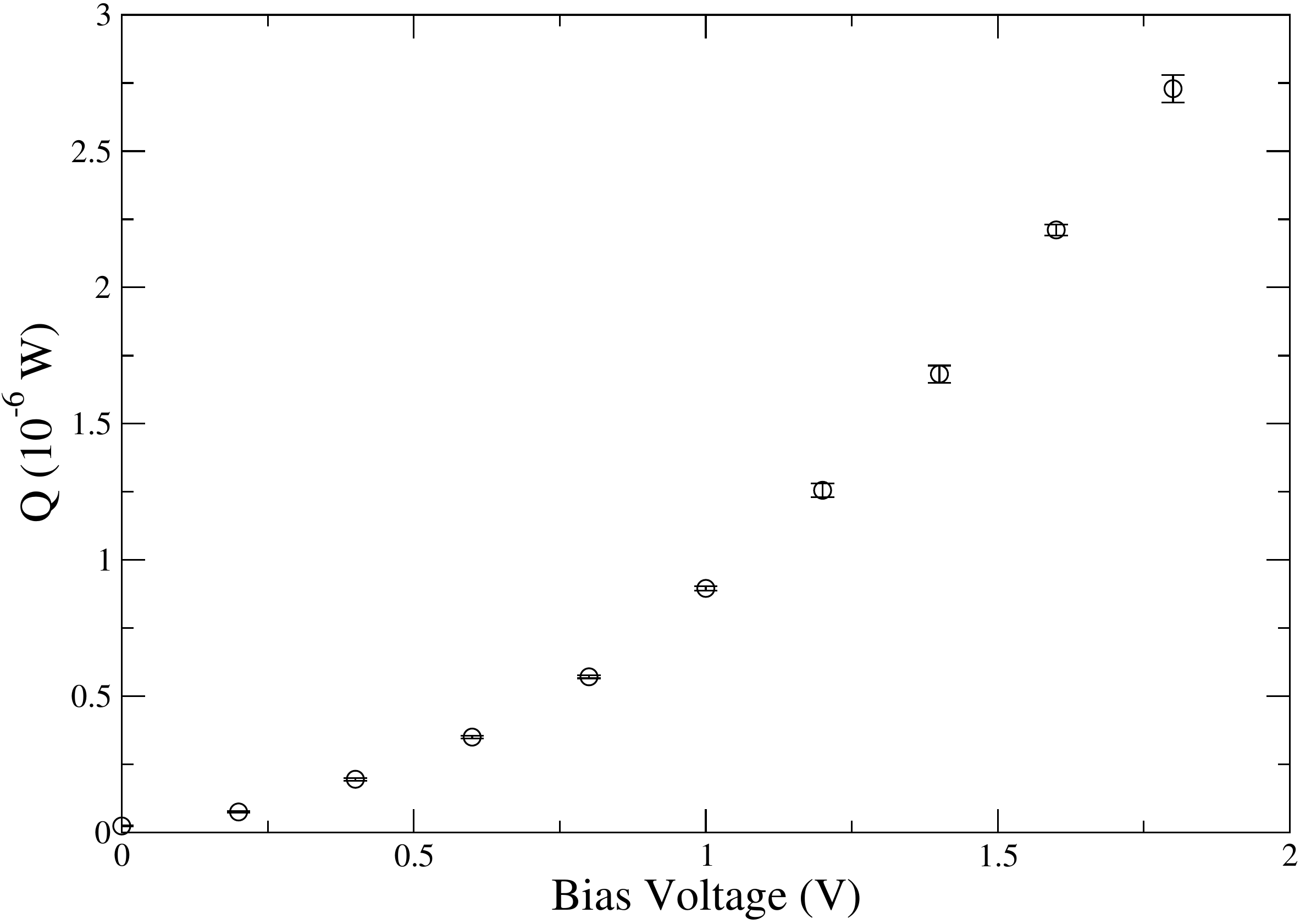}%
\caption{\label{fig-heatg}The heat generated per unit time versus
bias voltage of the electrons for a $4 \times 2$ graphene armchair configuration.}
\end{figure}

The rate of heat generation is plotted in Fig.~\ref{fig-heatg}.  The
result is for the same configuration as shown in Ref.~\onlinecite{WaNiJi.2009} of
Fig.~4.  Each data point takes about 4 days on a typical
Opteron CPU.  The error bars are quite small. The heat generation
at zero bias should be zero.  However, we get a small value.  This
has to do with the cut-off used in the noise for the electrons.  We have
used an abrupt cut-off for the spectrum at $\hbar \omega = 1.29\,$eV. 
The calculation demonstrates the feasibility of computing the Joule heating current, intrinsically a quantum effect at nanoscale, by classical molecular dynamics.

\section{Conclusions}
\label{sec:conc}
In summary, using a tight-binding-like Hamiltonian for the EPI, 
Eq.~(\ref{pheno-Htot}), we have introduced three different approaches to study
the effect of EPI in different parameter regimes. We focused on the electronic, phononic,
and thermoelectric transport properties of nano-conductors, in a general
multi-probe setup. For each approach, we started with the theoretical
derivation of the main equations. This was followed by applications in models or
simple systems, mainly for illustration purpose. 

Applications of these
approaches to more interesting problems are straightforward. Examples of such
problems are: (1) application of the NEGF and QME approach to nonequilibrium
thermoelectric transport to study the thermoelectric efficiency at finite
power output, (2) application of the QME approach to look at system where both
EPI and electron-electron interaction are important, (3) combining the
generalized Langevin equation with first-principles or tight-binding electronic
structure package to study current-induced dynamics in realistic
nano-conductors, especially to explore how the electron-dissipated heat is
transferred in and out of the nano-conductor. With these available tools, more
interesting and important systems can be investigated.

\begin{acknowledgments}
The authors thank M. Brandbyge, P. Hedeg{\aa}rd, Baowen Li for fruitful
collaborations and invaluable discussions. JTL and JWJ thank the hospitality of
Physics Department, National University of Singapore, where the paper is
finished. JTL is supported by the National Natural Science Foundation of China
(Grant No. 11304107 and 61371015). JWJ is supported by the Recruitment Program
of Global Youth Experts of China and the start-up funding from Shanghai
University.  J.-S. W acknowledges support of an FRC grant R-144-000-343-112.
\end{acknowledgments}

\bibliography{thermoelectric}
\end{document}